\documentclass{jfm}
\usepackage{graphicx}
\usepackage{epstopdf, epsfig}
\usepackage[normalem]{ulem}

\usepackage[abs]{overpic}
\usepackage{graphicx}
\usepackage{dcolumn}
\usepackage{bm}
\usepackage{placeins}
\usepackage{subfig}
\usepackage{amsmath}
\usepackage{color}
\usepackage{xcolor,cancel}

\shorttitle{Forward and backward in time pair-separation PDFs for inertial particles }
\shortauthor{A.D. Bragg}

\title{Analysis of the forward and backward in time pair-separation PDFs for inertial particles in isotropic turbulence}

\author{Andrew D. Bragg\aff{1}
  \corresp{\email{andrew.bragg@duke.edu}},
  }

\affiliation{\aff{1}Department of Civil and Environmental Engineering, Duke University, Durham, North Carolina, USA}

\begin{document}

\maketitle

\begin{abstract}
In this paper we investigate, using theory and Direct Numerical Simulations (DNS), the Forward In Time (FIT) and Backward In Time (BIT) Probability Density Functions (PDFs) of the separation of inertial particle-pairs in isotropic turbulence. In agreement with our earlier study (Bragg \emph{et al.}, Phys. Fluids \textbf{28}, 013305 (2016)), where we compared the FIT and BIT mean-square separations, we find that inertial particles separate much faster BIT than FIT, with the strength of the irreversibility depending upon the final/initial separation of the particle-pair and their Stokes number $St$. However, we also find that the irreversibility shows up in subtle ways in the behavior of the full PDF that it does not in the mean-square separation. In the theory, we derive new predictions, including a prediction for the BIT/FIT PDF for ${St\geq O(1)}$, and for final/initial separations in the dissipation regime. The prediction shows how caustics in the particle relative velocities in the dissipation range affect the scaling of the pair-separation PDF, leading to a PDF with an algebraically decaying tail. The predicted functional behavior of the PDFs is universal, in that it does not depend upon the level of intermittency in the underlying turbulence. We also analyze the pair-separation PDFs for fluid particles at short-times, and construct theoretical predictions using the multifractal formalism to describe the fluid relative velocity distributions. The theoretical and numerical results both suggest that the extreme events in the inertial particle-pair dispersion at the small-scales are dominated by their non-local interaction with the turbulent velocity field, rather than due to the strong dissipation range intermittency of the turbulence itself. In fact, our theoretical results predict that for final/initial separations in the dissipation range, when $St\gtrsim 1$, the tails of the pair-separation PDFs decay faster as $Re_\lambda$ is increased, the opposite of what would be expected for fluid particles.
\end{abstract}

\begin{keywords}

\end{keywords}

\section{Introduction}

The dispersion of particles in turbulent flows is a problem of great fundamental interest, with a wide range of applications in industrial contexts \citep{fox12}, cloud physics \citep{devenish12}, astrophysics \citep{pan10}, plankton distribution in oceans \citep{delillo14}, and the dispersion of plant seeds \citep{heydel2014}, to name but a few. An aspect of particular importance is understanding how particles in these systems move relative to each other as a function of time, which, for example, can be used to quantify the efficiency of turbulent mixing \citep{sawford05}. In this paper, we shall be concerned with pair-dispersion, where the separation between two particles in a turbulent flow is considered as a function of time.

Dynamically, the simplest kind of particle-pair dispersion one can consider is that of fluid particles (tracers) that precisely follow the local turbulent fluid velocity field. Even this problem is, however, very difficult because of the complexity of the turbulence itself. This topic has been the subject of intense investigation, with some of the original pioneering work by \citet{batchelor52b}, \citet{richardson,richardson26}, continuing right up to more recent times, including theoretical, numerical and experimental studies \citep{falkovich01,boffetta02a,boffetta02b,biferale05,ouellette06c,salazar09,bec10b,ni13,biferale14,buaria15,bragg16}. In many applications, however, the particles cannot be considered simple fluid particles, and they often posses non-negligible inertia. The challenge then is to understand and predict how inertia affects the turbulent pair-separation   \citep{fouxon08,bec10b,biferale14,bragg16}. 

Irrespective of the kinds of particles considered, the majority of pair-separation studies have focused on Forward In Time (FIT) dispersion, where the \emph{initial} separation of the particle-pair is chosen, and their separation at \emph{later} times is analyzed. This type of dispersion is important for understanding how groups of particles spread out in turbulent flows, such as the evolution of plumes of ash particles emitted during volcanic eruptions    \citep{folch12}. The other kind of dispersion is Backward In Time (BIT) dispersion, where the \emph{final} separation of the particle-pair is chosen, and their separation at \emph{earlier} times is analyzed. This type of dispersion is important for mixing problems, for which the rate of BIT dispersion quantifies how efficiently the particles are mixing in the system. For inertial particles, BIT dispersion is also crucial for understanding and predicting their relative velocities \citep{pan10} and clustering, since their clustering is directly connected to the behavior of their relative velocities \citep{bragg14b}.

An important question concerns the irreversibility of the dispersion, quantified by the difference between FIT and BIT dispersion. The FIT and BIT dispersion of fluid particles has been considered in a number of studies \citep{sawford05,berg06a,falkovich13,jucha14,xu15,buaria15,buaria16,pumir16}, with the conclusion that in 3D turbulence, fluid particle-pairs separate faster BIT than FIT. The effect of particle inertia on the BIT dispersion was only recently addressed by \citet{bragg16}. They found that inertia has a profound effect upon the dispersion irreversibility, with inertial particle dispersion being much more strongly irreversible than fluid particle dispersion, in general. However, the analysis by \citet{bragg16} only considered the mean-square separation of the inertial particles. To fully understand and characterize the BIT dispersion and the dispersion irreversibility, the full Probability Distribution Functions (PDFs) of the particle-pair separations must be analyzed FIT and BIT. This is precisely the purpose of the present paper. The FIT pair-separation PDFs for inertial particles have already been investigated by \citet{bec10b} and \citet{biferale14}, and we will therefore focus more on the BIT PDF and the irreversibility of the pair-separation.

The outline of the rest of the paper is as follows. In \S\ref{theory} we examine theoretically the FIT and BIT pair-separation PDFs, considering the effect of irreversibility mechanisms, and we derive new predictions. Then in \S\ref{RaD} we use data from particle-laden Direct Numerical Simulations (DNS) to further analyze the dispersion PDFs, and test the theoretical predictions.

\section{Theory}\label{theory}
\subsection{Irreversibility}\label{IM}
Following on from the work of \citet{bragg16}, we shall consider statistically stationary, homogeneous, isotropic turbulence, in which the inertial particle statistics are also stationary. From a practical perspective, this is perhaps not the most relevant system in which to study the irreversibility of particle-pair dispersion. For example, in many practical dispersion problems, particles will often be injected into the turbulent flow with velocities such that the as the particles begin to disperse, they will undergo an initial transitory phase before their single-time statistics become stationary. However, from a fundamental perspective, a statistically stationary system is the most natural starting point for examining irreversibility, since in such a system, trivial causes of irreversibility (e.g. global energy decay/growth) are removed, and the mechanisms generating irreversibility must arise solely from the intrinsic dynamics of the system. In future work we will consider non-stationary effects, in order to assess more carefully the implications of irreversible dispersion for practical problems.  

For the system of interest, the statistics of the pair-separation only depend upon the evolution of their separation vector $\bm{r}^p$ and not their center of mass. The FIT and BIT PDFs describing the particle-pair dispersion process may then be defined as
\begin{align}
\mathcal{P^F}(\bm{r},t\vert\bm{\xi},t')&\equiv\Big\langle \delta\Big(\bm{r}^p(t)-\bm{r}\Big)\Big\rangle_{\bm{r}^p(t')=\bm{\xi}},\,\quad t'\in[0,t],\label{Pf}\\
\mathcal{P^B}(\bm{r},t'\vert\bm{\xi},t)&\equiv\Big\langle \delta\Big(\bm{r}^p(t')-\bm{r}\Big)\Big\rangle_{\bm{r}^p(t)=\bm{\xi}},\quad t'\in[0,t],\label{Pb}
\end{align}
where $\bm{r}$ denotes the time-independent configuration space coordinate, and $\langle\cdot\rangle_{\bm{r}^p(t')=\bm{\xi}}$ denotes an ensemble average conditioned upon $\bm{r}^p(t')=\bm{\xi}$, and similarly for the BIT case.
Due to isotropy, we need only consider the PDFs for $\|\bm{r}^p\|$, that is
\begin{align}
\mathcal{P^F}({r},t\vert{\xi},t')&\equiv\Big\langle \delta\Big(\|\bm{r}^p(t)\|-{r}\Big)\Big\rangle_{\|\bm{r}^p(t')\|={\xi}},\,\quad t'\in[0,t],\label{PDFf}\\
\mathcal{P^B}({r},t'\vert{\xi},t)&\equiv\Big\langle \delta\Big(\|\bm{r}^p(t')\|-{r}\Big)\Big\rangle_{\|\bm{r}^p(t)\|={\xi}},\quad t'\in[0,t],\label{PDFb}
\end{align}
where $\xi\equiv \|\bm{\xi}\|$. One way to analyze the behavior of these PDFs is to consider their evolution equations $\partial_t\mathcal{P^F}$ and $\partial_{t'}\mathcal{P^B}$. We find it more insightful, however, to consider integral formulations for the PDFs that may be constructed by substituting into (\ref{PDFf}) and (\ref{PDFb}) the integral representations of the solutions for $\bm{r}^p(t)$ and $\bm{r}^p(t')$, respectively.  

From the kinematic equation $\dot{\bm{r}}^p(t)\equiv\bm{w}^p(t)$ we obtain
\begin{align}
d_t\|\bm{r}^p(t)\|^2\equiv 2\|\bm{r}^p(t)\| w^p_\parallel(t),
\end{align}
where $w^p_\parallel(t)\equiv\|\bm{r}^p(t)\|^{-1}\bm{r}^p(t)\bm{\cdot}\bm{w}^p(t) $, leading to the result
\begin{align}
\|\bm{r}^p(t)\|\equiv \Bigg[\|\bm{r}^p(t')\|^2+2\int_{t'}^t\|\bm{r}^p(t'')\| w^p_\parallel(t'')\,dt''\Bigg]^{1/2}.
\end{align}
Using this result in (\ref{PDFf}) and (\ref{PDFb}) we obtain
\begin{align}
\mathcal{P^F}({r},t\vert{\xi},t')&=\Bigg\langle \delta\Bigg(\Big[\xi^2+2\int_{t'}^t\|\bm{r}^p(t'')\| w^p_\parallel(t'')\,dt''\Big]^{1/2}-{r}\Bigg)\Bigg\rangle_{\|\bm{r}^p(t')\|={\xi}},\label{PDFf2}\\
\mathcal{P^B}({r},t'\vert{\xi},t)&=\Bigg\langle \delta\Bigg(\Big[\xi^2-2\int_{t'}^t\|\bm{r}^p(t'')\| w^p_\parallel(t'')\,dt''\Big]^{1/2}-{r}\Bigg)\Bigg\rangle_{\|\bm{r}^p(t)\|={\xi}}.\label{PDFb2}
\end{align}
Since we are interested in the statistically stationary state of the system, for which the dispersion only depends upon the time difference $s\equiv t-t'$, we may set the conditioning time to zero, and note the time ordering $t'\leq t$, to obtain
\begin{align}
\mathcal{P^F}({r},s\vert{\xi},0)&=\Bigg\langle \delta\Bigg(\Big[\xi^2+2\int_{0}^s\|\bm{r}^p(s')\| w^p_\parallel(s')\,ds'\Big]^{1/2}-{r}\Bigg)\Bigg\rangle_{{\xi}},\label{PDFf3}\\
\mathcal{P^B}({r},-s\vert{\xi},0)&=\Bigg\langle \delta\Bigg(\Big[\xi^2-2\int_{-s}^0\|\bm{r}^p(s')\| w^p_\parallel(s')\,ds'\Big]^{1/2}-{r}\Bigg)\Bigg\rangle_{{\xi}},\label{PDFb3}
\end{align}
where $s\geq 0$, and $\langle\cdot\rangle_\xi\equiv \langle\cdot\rangle_{\|\bm{r}^p(0)\|={\xi}}$. Note that, to be precise, whereas in the FIT case, $s\in[0,\infty]$, in the BIT case $s\in[0,T_0]$, where $-s=-T_0$ corresponds to the initial time of the dynamical system, since in the original variables $t'\in[0,t]$ not $t'\in[-\infty,t]$. However, since we are considering the stationary state, we could effectively set $T_0\to\infty$.

In \citet{bragg16} we presented arguments for the physical mechanisms that generate irreversible pair-dispersion in turbulence. Here we will present the arguments in alternative form, especially to bring out in more detail some aspects of the problem that were not considered thoroughly in \citet{bragg16}.

There are essentially two ways that (\ref{PDFf3}) and (\ref{PDFb3}) could differ. First, they will differ trivially if the statistics of $w^p_\parallel$ (conditioned on $\|\bm{r}^p(0)\|=\xi$) differ for $s$ and $-s$. In one sense this effect would appear to be absent for the system we are considering because of statistical stationarity. However, the conditioning $\langle\cdot\rangle_\xi$ actually means that (\ref{PDFf3}) and (\ref{PDFb3}) depend upon multi-time statistics of $w^p_\parallel$, and as such they could differ for $s$ and $-s$. The second effect is non-trivial: Equation (\ref{PDFf3}) reveals that only particle-pairs whose motion is dominated by $w^p_\parallel>0$ (i.e. such that $\int_{0}^s\|\bm{r}^p(s')\| w^p_\parallel(s')\,ds'>0$) will go to larger separations $r>\xi$, whereas only particle-pairs whose motion is dominated by $w^p_\parallel<0$ (i.e. such that $\int_{0}^s\|\bm{r}^p(s')\| w^p_\parallel(s')\,ds'<0$) will go to smaller separations $r<\xi$. Conversely, the BIT result in (\ref{PDFb3}) reveals that only particle-pairs whose motion was dominated by $w^p_\parallel<0$ were at larger separations $r>\xi$ in the past, whereas only particle-pairs whose motion was dominated by $w^p_\parallel>0$ were at smaller separations $r<\xi$ in the past. It follows then that if the PDFs of $w^p_\parallel$ are themselves asymmetric, the separation PDF will be irreversible\footnote[1]{This is actually a necessary but not sufficient criteria for dispersion irreversibility. The other condition required is that the correlation timescales of $w^p_\parallel$ must be finite, but this condition is always satisfied in real turbulent flows. See \citet{bragg16} for further details.}, i.e. $\mathcal{P^F}({r},s\vert{\xi},0)\neq \mathcal{P^B}({r},-s\vert{\xi},0)$.

These observations are purely kinematic; the dynamical origin of any irreversibility depends upon the dynamics governing $\bm{w}^p$. Following on from the work in \citet{bragg16}, we shall consider heavy, point particles whose motion is governed by a Stokes drag force. In this case
\begin{align}
\dot{\bm{w}}^p(t)=\frac{1}{\tau_p}\Big(\Delta\bm{u}(\bm{x}^p(t),\bm{r}^p(t),t)-\bm{w}^p(t)\Big),\label{eom1}
\end{align}
where $\bm{x}^p(t)$ and $\bm{x}^p(t)+\bm{r}^p(t)$ are the locations of the two particles, $\bm{w}^p(t)$ is their relative velocity, and $\Delta\bm{u}(\bm{x}^p(t),\bm{r}^p(t),t)$ is the difference in the fluid velocity evaluated at the two particle positions. To avoid any confusion, we note that in (\ref{eom1}), the time label ``$t$'' is generic and does not necessarily coincide with the ``$t$'' appearing in the expressions (\ref{PDFf}) and (\ref{PDFb}). Furthermore, since we are considering homogeneous turbulence, we will usually neglect the $\bm{x}^p(t)$ argument when discussing statistical properties of the dispersion.

In the absence of turbulence (i.e. setting $\Delta\bm{u}=\bm{0}$), (\ref{eom1}) would reduce to $\dot{\bm{w}}^p(t)=-\bm{w}^p(t)/\tau_p$, and the dispersion would be trivially irreversible simply because of dissipation, giving $d_t\|\bm{w}^p(t)\|^2<0$ (where $d_t\equiv d/dt$). However, in the presence of turbulence, $d_t\|\bm{w}^p(t)\|^2$ is not strictly negative since the fluid does work on the particles. In this case, understanding the irreversibility is much more complex since the behavior of $\bm{w}^p$ then depends upon how the inertial particle-pairs interact with the field $\Delta\bm{u}$, and the properties of $\Delta\bm{u}$ itself. More specifically, since we are interested in statistical irreversibility, we must understand how $w^p_\parallel$ in (\ref{PDFf3}) and (\ref{PDFb3}) depends upon the statistical properties of $\Delta\bm{u}$. A crucial point in this regard is that in (\ref{PDFf3}) and (\ref{PDFb3}), the behavior of $w^p_\parallel(s')$ is not only controlled by its dynamical equation, but also by the fact that the trajectories (along which $w^p_\parallel(s')$ is measured) that contribute to the ensembles in (\ref{PDFf3}) and (\ref{PDFb3}) \emph{are conditioned}, i.e. the contributing trajectories have to satisfy ${\|\bm{r}^p(s'=0)\|=\xi}$. This trajectory conditioning is the reason why FIT and BIT separating pairs experience values of $w^p_\parallel(s')$ with different signs, as discussed earlier.

The forward solution to (\ref{eom1}), satisfying the condition ${\|\bm{r}^p(s'=0)\|=\xi}$, may be written as
\begin{align}
\bm{w}^p(s'\vert\xi,0)=e^{-s'/\tau_p}\bm{w}^p(0\vert\xi,0)+\frac{1}{\tau_p}\int\limits^{s'}_0 e^{-(s'-s'')/\tau_p}\Delta\bm{u}(\bm{r}^p(s''\vert\xi,0),s'')\,ds''.\label{wpsolF}
\end{align}
The backward solution for $\bm{w}^p$ can be represented by (\ref{wpsolF}) with $s'\to -s'$ and $s'\in[0,T_0]$, in which case the backward trajectory is represented as a terminal value problem. In this time-reversed solution, the term $e^{s'/\tau_p}\bm{w}^p(0\vert\xi,0)$ does not cause the solution $\bm{w}^p(-s'\vert\xi,0)$ to blow up. This is because $\bm{w}^p(0\vert\xi,0)$ is not an arbitrary end condition for the solution (unlike the forward solution where $\bm{w}^p(0\vert\xi,0)$ can be chosen arbitrarily), but implicitly contains the information on the trajectory history of the particle, and this regularizes the solution to the time-reversed equation. This can be seen by constructing $\bm{w}^p(-s'\vert\xi,0)$ as an initial value solution, with initial time $-s'=-T_0$
\begin{align}
\bm{w}^p(-s'\vert\xi,0)=e^{(s'-T_0)/\tau_p}\bm{w}^p(-T_0\vert\xi,0)+\frac{1}{\tau_p}\int\limits^{-s'}_{-T_0} e^{(s'+s'')/\tau_p}\Delta\bm{u}(\bm{r}^p(s''\vert\xi,0),s'')\,ds''.\label{wpsolB}
\end{align}
Setting $s'=0$ in this solution shows how $\bm{w}^p(0\vert\xi,0)$ depends upon the trajectory history of the particle, and by inserting this expression for $\bm{w}^p(0\vert\xi,0)$ into the time-reversed form of (\ref{wpsolF}), one indeed finds that the resulting solution does not blow up. We will find both the ``terminal value' and ``initial value'' representations of the backward solution $\bm{w}^p(-s'\vert\xi,0)$ useful in our discussion.

If $\bm{w}^p(0\vert\xi,0)\neq\bm{0}$, then in the regime $s'/\tau_p\ll 1$ we have
\begin{align}
\Big\|e^{-s'/\tau_p}\bm{w}^p(0\vert\xi,0)\Big\|\gg\Big\|\frac{1}{\tau_p}\int\limits^{s'}_0 e^{-(s'-s'')/\tau_p}\Delta\bm{u}(\bm{r}^p(s''\vert\xi,0),s'')\,ds''\Big\|,
\end{align}
which shows that in this regime, the evolution of $\bm{w}^p(s'\vert\xi,0)$ is determined by dissipation. Inserting the leading order behavior $\bm{w}^p(s'\vert\xi,0)\approx e^{-s'/\tau_p}\bm{w}^p(0\vert\xi,0)$ into (\ref{PDFf3}) and (\ref{PDFb3}) leads to $\mathcal{P^F}({r},s\vert{\xi},0)\neq \mathcal{P^B}({r},-s\vert{\xi},0)$. In this regime the dispersion is irreversible simply because of dissipation, and particles will separate faster BIT than FIT simply because the particle kinetic energy is decaying in time. Formally, this dissipation effect will dominate the irreversibility in the short-time regime $s/\tau_p\ll1$. However, its effect can persist up to $s/\tau_p\lesssim O(1)$ in certain regimes of the flow, a point we shall discuss further in \S\ref{TheorScal}. In \citet{bragg16} we did not properly consider this short-time, dissipation-induced irreversibility mechanism, and as we shall now show, the irreversibility mechanisms identified in \citet{bragg16} actually apply when $s/\tau_p\not\ll1$.

After the initial transient stage, i.e. for  $s'/\tau_p>1$, the forward and backward solutions for $\bm{w}^p$ become (using (\ref{wpsolB}) for the backward case)
\begin{align}
\bm{w}^p(s'\vert\xi,0)&\approx\frac{1}{\tau_p}\int\limits^{s'}_0 e^{-(s'-s'')/\tau_p}\Delta\bm{u}(\bm{r}^p(s''\vert\xi,0),s'')\,ds'',\label{wpsolFb}\\
\bm{w}^p(-s'\vert\xi,0)&\approx\frac{1}{\tau_p}\int\limits^{-s'}_{-T_0} e^{(s'+s'')/\tau_p}\Delta\bm{u}(\bm{r}^p(s''\vert\xi,0),s'')\,ds''.\label{wpsolBb}
\end{align}
As discussed earlier, if the PDFs of $\bm{w}^p$ are asymmetric, then  $\mathcal{P^F}({r},s\vert{\xi},0)\neq \mathcal{P^B}({r},-s\vert{\xi},0)$. In \citet{bragg16}, by considering the behavior of $\bm{w}^p(s'\vert\xi,0)$ and $\bm{w}^p(-s'\vert\xi,0)$ we argued that there are two distinct physical mechanisms that generate such asymmetry in the PDFs of $\bm{w}^p$. We refer the reader to that paper for detailed explanations; here we summarize. Let us first define $St_r(s')\equiv\tau_p/\tau_r(s')$, where $\tau_r(s')$ is the eddy turnover time based upon the particle separation at time $s'$. If we project (\ref{wpsolFb}) and (\ref{wpsolBb}) onto $\bm{r}^p(s'\vert\xi,0)$ and $\bm{r}^p(-s'\vert\xi,0)$, respectively, then in the regime $St_r(s')\ll 1$ we obtain
\begin{align}
{w}_\parallel^p(s'\vert\xi,0)&\approx\Delta{u}_\parallel(\bm{r}^p(s'\vert\xi,0),s'),\label{wpsolFc}\\
{w}_\parallel^p(-s'\vert\xi,0)&\approx\Delta{u}_\parallel(\bm{r}^p(-s'\vert\xi,0),-s').\label{wpsolBc}
\end{align}
As noted earlier, because of the trajectory conditioning, particles that are separating FIT at time $s'$ have ${w}_\parallel^p(s'\vert\xi,0)>0$, whereas particles that are separating BIT at time $-s'$ have ${w}_\parallel^p(-s'\vert\xi,0)<0$. However, due to the dynamical fluxes in the turbulent velocity field, the PDF of $\Delta u_\parallel$ is negatively skewed in 3D. Consequently, we expect (statistically) that $\vert{w}_\parallel^p(-s'\vert\xi,0)\vert>\vert{w}_\parallel^p(s'\vert\xi,0)\vert$, i.e. the energy flux in 3D turbulence causes particle-pairs with $St_r\ll 1$ to be pushed together more strongly than apart. In \citet{bragg16} we referred to this as the Local Irreversibility Mechanism (LIM), since it arises from the local (in a temporal sense) properties of $\Delta u_\parallel$ experienced by the particle-pair. For fluid particles, the LIM is the only irreversibility mechanism since they do not experience the short-time dissipation source of irreversibility (because they do not experience a drag force) described earlier.

When $St_r(s')\not\ll 1$, the path-history contribution in (\ref{wpsolFb}) and (\ref{wpsolBb}) is important, and the inertial particle dynamics is temporally non-local. Since particles that are separating FIT at time $s'$ have ${w}_\parallel^p(s'\vert\xi,0)>0$, then their separation would have been \emph{smaller} in the past. On the other hand, since particles that are separating BIT at time $-s'$ have ${w}_\parallel^p(-s'\vert\xi,0)<0$, then their separation would have been \emph{larger} in the past. Since $\Delta\bm{u}$ on average increases with increasing separation (true instantaneously in the dissipation range) then the path-integral in (\ref{wpsolBb}) will be (statistically speaking) larger than that in (\ref{wpsolFb}). Consequently, we expect (statistically) that $\vert{w}_\parallel^p(-s'\vert\xi,0)\vert>\vert{w}_\parallel^p(s'\vert\xi,0)\vert$, just as for the LIM case. In \citet{bragg16} this was referred to as the Non-Local Irreversibility Mechanism (NLIM), since it arises from the non-local dynamics of the inertial particles, and operates when $St_r(s')\not\ll 1$ provided that the particle separation lies in the range where the statistics of $\Delta\bm{u}$ depend upon separation (i.e. at sub-integral scales). As emphasized in \citet{bragg16}, the NLIM \emph{does not} arise from skewness in the underlying field $\Delta\bm{u}$, and would operate even in a kinematic field where $\Delta\bm{u}$ has a symmetric PDF.

Whereas in the regime $s\ll\tau_p$, dissipation plays an explicit role in the irreversibility, its effect is implicit outside of this regime. Dissipation is implicit in the NLIM since the memory of the inertial particles (on which the NLIM depends) comes from their dissipative dynamics. However, dissipation is not the actual cause of irreversibility outside the regime $s\ll\tau_p$. This is demonstrated by the following argument: If the statistics of the field $\Delta\bm{u}$ were independent of separation, then the system described by (\ref{eom1}) would be mathematically identical to the case of single inertial particles moving in a homogeneous turbulent flow field, in which the particle velocity PDFs are necessarily symmetric. For such a system, the statistics of (\ref{wpsolFb}) and (\ref{wpsolBb}) would be identical, and through (\ref{PDFf3}) and (\ref{PDFb3}) this would imply reversible dispersion. This is consistent with the known fact that the statistics of single-particle velocities in stationary, homogeneous turbulence are reversible \citep{falkovich12}. Thus, dissipation alone cannot break the symmetry of the PDF of $\bm{w}^p$ and therefore it is not, in and of itself, the cause of irreversibility outside the regime $s\ll\tau_p$.

Another important point is that the asymmetry of the PDF of $\bm{w}^p$ also plays a role in the regime $s\ll\tau_p$, where the explicit source of irreversibility comes from the particle dissipation. In the regime $s\ll\tau_p$ we have the FIT behavior ${w}_\parallel^p(s'\vert\xi,0)\approx e^{-s'/\tau_p}{w}_\parallel^p(0\vert\xi,0)$, and BIT we have ${w}_\parallel^p(-s'\vert\xi,0)\approx e^{s'/\tau_p}{w}_\parallel^p(0\vert\xi,0)$. However, since FIT separating particles must have ${w}_\parallel^p(s'\vert\xi,0)>0$, and BIT have ${w}_\parallel^p(-s'\vert\xi,0)<0$, then even in the short-time limit, the irreversibility will be affected by the asymmetry of the PDF of ${w}_\parallel^p(0\vert\xi,0)$.

In addition to the role of dissipation in the short-time limit, another issue that was not thoroughly considered in \citet{bragg16} is the role of ``preferential sampling'' on the irreversibility. Preferential sampling relates to the fact that due to the way inertial particles interact with the topology of the fluid velocity field, they do not uniformly sample the field $\Delta\bm{u}(\bm{r},t)$, showing a tendency to avoid rotation dominated regions of the flow \citep{maxey87,ireland16a}. This effect is implicit in both LIM and NLIM, because in either case, the fluid velocity difference driving the particle relative motion is $\Delta\bm{u}(\bm{r}^p(t),t)$, the statistics of which (conditioned on ${\bm{r}^p(t)=\bm{r}}$) will deviate from those of $\Delta\bm{u}(\bm{r},t)$ due to preferential sampling. Both the LIM and the NLIM would still generate irreversible dispersion even in the absence of preferential sampling, however, their strength will be quantitatively affected by preferential sampling. For example, in the local case, the strength of the dispersion irreversibility will be affected since the asymmetry in the PDF of $\Delta\bm{u}(\bm{r}^p(t),t)$ may not be the same as the asymmetry in the PDF of $\Delta\bm{u}(\bm{r},t)$. Similarly, in the non-local case, the strength of the dispersion irreversibility will be affected by the fact that the values of $\Delta\bm{u}(\bm{r}^p(t),t)$ along the path-history of the particle pair will be biased due to preferential sampling \citep[similar to how the non-local clustering mechanism in the dissipation range is affected by the preferential sampling effect, see][]{bragg14d}. The only assumption we are making is that the skewness of the PDF of $\Delta\bm{u}(\bm{r}^p(t),t)$ (more precisely, its longitudinal component) remains negative in 3D $\forall St$, and that the moments of $\Delta\bm{u}(\bm{r}^p(t),t)$ increase with increasing separation $\forall St$. In other words, we assume that preferential sampling causes the statistics of $\Delta\bm{u}(\bm{r}^p(t),t)$ to differ from those of $\Delta\bm{u}(\bm{r},t)$ quantitatively, but not qualitatively. These assumptions seem very reasonable, and will be validated using DNS data in \S\ref{RaD}.

In \citet{bragg16}, we used the arguments for the irreversibility mechanisms to predict that the mean-square particle separation should be faster BIT than FIT $\forall St_r$, which was confirmed by DNS results. When considering the full PDFs $\mathcal{P^F}({r},s\vert{\xi},0)$ and $\mathcal{P^B}({r},-s\vert{\xi},0)$, the effect is more subtle. In particular, we must consider the behavior for $r<\xi$ and $r>\xi$ separately. 

As discussed earlier, for $r>\xi$, $\mathcal{P^F}({r},s\vert{\xi},0)$ is governed by particles whose motion is dominated by $w^p_\parallel>0$, whereas for $r>\xi$, $\mathcal{P^B}({r},-s\vert{\xi},0)$ is governed by particles whose motion was dominated by $w^p_\parallel<0$. Since the PDFs for $w^p_\parallel$ are negatively skewed then we would expect that for $r>\xi$, $\mathcal{P^B}({r},-s\vert{\xi},0)>\mathcal{P^F}({r},s\vert{\xi},0)$. However, for $r<\xi$ the relationship is inverted: For $r<\xi$, $\mathcal{P^F}({r},s\vert{\xi},0)$ is governed by particles whose motion is dominated by $w^p_\parallel<0$, whereas for $r<\xi$, $\mathcal{P^B}({r},-s\vert{\xi},0)$ is governed by particles whose motion was dominated by $w^p_\parallel>0$. Since the PDFs for $w^p_\parallel$ are negatively skewed then we would expect that for $r<\xi$, $\mathcal{P^F}({r},s\vert{\xi},0)>\mathcal{P^B}({r},-s\vert{\xi},0)$.

With respect to this prediction, three remarks are in order. First, $r>\xi$ and $r<\xi$ in the previous argument should not be taken too literally, since we would expect a transition region between the two predicted behaviors. Second, the prediction that for $r>\xi$, $\mathcal{P^B}({r},-s\vert{\xi},0)>\mathcal{P^F}({r},s\vert{\xi},0)$ is consistent with the findings in \citet{bragg16} for the mean-square separations, namely that the BIT mean-square separation should be greater than the FIT counterpart. The connection is that the mean-square separation is itself dominated by the behavior of particles at $r>\xi$, i.e. $\langle\|\bm{r}^p(\pm s)\|^2\rangle_{\xi}\geq\xi^2$. Third, for $\xi$ sufficiently small and/or $s$ sufficiently large, the prediction for $r<\xi$ may not hold due to the effect of particle-pairs whose separation $\|\bm{r}^p\|$ has passed through a minimum on the interval $s'\in[0,s]$.

So far we have considered how, for a given $St$, $\mathcal{P^F}$ and $\mathcal{P^B}$ might differ. Another important question is how $\mathcal{P^F}$ and $\mathcal{P^B}$ are each affected by changes in $St$. To understand this, we consider the relative velocities of inertial particle governed by (\ref{eom1}) at a given separation $\bm{r}$ (ignoring the initial condition and the $\bm{x}$ coordinate)
\begin{align}
\bm{w}^p(t\vert\bm{r})=\frac{1}{\tau_p}\int\limits^{t}_0 e^{-(t-s)/\tau_p}\Delta\bm{u}(\bm{r}^p(s\vert\bm{r},t),s)\,ds,\label{wpsol}
\end{align}
where $\bm{w}^p(t\vert\bm{r})$ denotes the value of $\bm{w}^p$ at time $t$ conditioned upon $\bm{r}^p(t)=\bm{r}$. The solution (\ref{wpsol}) may be split up into its local and non-local contributions \footnote[2]{By local and non-local we are referring to the way $\bm{w}^p$ depends dynamically upon the field $\Delta\bm{u}(\bm{r},t)$. There is another kinematic kind of non-locality: $\Delta\bm{u}(\bm{r}^p(t\vert\bm{r},t),t)$ is kinematically non-local since $\bm{r}^p(t\vert\bm{r},t)$ itself depends upon a time integral of $\bm{w}^p$ because $\dot{\bm{r}}^p(t)\equiv\bm{w}^p(t)$. In this kinematic sense, even fluid particle dynamics would be non-local (their dynamics is also non-local due to the non-local pressure in the Navier-Stokes equation, however this is a spatial non-locality, not temporal, which is the kind we are discussing here). In all of our discussion, by local and non-local we only refer to the temporal dependence of $\bm{w}^p$ on $\Delta\bm{u}$, which is only non-local for inertial, not fluid particles.}. The local contribution is simply the part that comes from the replacement $\Delta\bm{u}(\bm{r}^p(s\vert\bm{r},t),s)\to\Delta\bm{u}(\bm{r}^p(t\vert\bm{r},t),t)$ in (\ref{wpsol}), allowing us to write
\begin{align}
\begin{split}
\bm{w}^p(t\vert\bm{r})=&\underbrace{\Delta\bm{u}(\bm{r}^p(t\vert\bm{r},t),t)}_{\text{Local}}+\underbrace{\frac{1}{\tau_p}\int\limits^{t}_0 e^{-(t-s)/\tau_p}\Big(\Delta\bm{u}(\bm{r}^p(s\vert\bm{r},t),s)-\Delta\bm{u}(\bm{r}^p(t\vert\bm{r},t),t)\Big)\,ds}_{\text{Non-Local}}.
\end{split}\label{wpsol2}
\end{align}
The local contribution is the same, regardless of whether the pair-separation was larger or smaller in the past, but not so for the non-local contribution. Similar to the arguments for the non-local irreversibility mechanism, the non-local contribution in (\ref{wpsol2}) will be larger for particles whose separation satisfies $\|\bm{r}^p(s\vert\bm{r},t)\|>{r}$ than for those satisfying $\|\bm{r}^p(s\vert\bm{r},t)\|<{r}$ for $s<t$. The local contribution in (\ref{wpsol2}) will therefore (statistically) be more significant for separating particles than approaching particles, and in this sense the former is less influenced by the effects of particle inertia than the latter (noting that only the local part survives in the limit of weak inertia). Coupled with the previous irreversibility arguments, this implies that for $r>\xi$, $\mathcal{P^B}$ should be more strongly affected by particle inertia than $\mathcal{P^F}$, but the opposite for $r<\xi$. 
\subsection{Dependence of the PDFs on $s,\xi,r$}\label{TheorScal}
We now turn to consider how $\mathcal{P^F}$ and $\mathcal{P^B}$ depend on $s,\xi,r$. An important quantity in this respect is the timescale $\tau^p_\xi$, defined as the timescale corresponding to the autocovariance $\langle w^p_\parallel(0)w^p_\parallel(s')\rangle_{\xi}$.

The first regime to consider is the ``short-time regime''. Although formally this regime is $s/\tau^p_\xi\ll1$, theoretical results based on this asymptotic limit are often accurate up to $s/\tau^p_\xi\leq O(1)$. From a physical perspective, the significance of this regime is that it corresponds to the regime where the dispersion behavior is dominated by the particle relative velocities at the conditioning time $s=0$. 

We begin by considering the regime $\xi\ll\eta$ and $St\geq O(1)$, for which a question of significant interest pertains to understanding how the presence of caustics in the inertial particle relative velocities in the dissipation range \citep{wilkinson05,wilkinson06,gustavsson11} influences the behavior of $\mathcal{P^B}$. As discussed earlier, in the short-time regime $s\ll\tau_p$ we have $\bm{w}^p(-s')\approx e^{s'/\tau_p}\bm{w}^p(0)$, and from this we obtain
\begin{align}
\bm{r}^p(-s)&\approx\bm{r}^p(0)+\mathcal{G}(-s)\bm{w}^p(0),\label{rpsol}
\end{align}
where $\mathcal{G}(-s)\equiv\tau_p(1-e^{s/\tau_p})$. This approximation is valid provided that
\begin{align}
\Big\| e^{s/\tau_p}\bm{w}^p(0)\Big\|\gg\Big\|\frac{1}{\tau_p}\int^0_{-s} e^{(s'+s'')/\tau_p}\Delta\bm{u}(\bm{r}^p(s'),s')\,ds'\Big\|.\label{CA}
\end{align}
Although this regime is typically only realized for $s\ll\tau_p$, in the caustic regions it may be valid for much longer times, say for $s\leq O(\tau_p)$, because in the caustic regions $\|\bm{w}^p\|\gg\|\Delta\bm{u}\|$. As such, (\ref{rpsol}) may be valid beyond the ``short-time'' regime $s\ll\tau_p$, especially when $\bm{w}^p(0)$ corresponds to the large values described by the heavy tails of its PDF, tails that are much heavier than those for the PDF of $\Delta\bm{u}$ (see \citet{bec10a,ireland16a}). 

Using (\ref{rpsol}) we obtain
\begin{align}
\mathcal{P^B}({r},-s\vert{\xi},0)\approx\Big\langle \delta\Big(\|\bm{\xi}+\mathcal{G}(-s)\bm{w}^p(0)\|-{r}\Big)\Big\rangle_{{\xi}}.
\end{align}
Although only formally valid for the regime $St\gg1$, DNS results show that for particle-pairs in the dissipation range with $St\geq O(1)$, the caustics in their relative velocities lead to the statistics of the longitudinal and perpendicular projections of $\bm{w}^p(0)$ being approximately equal \citep{ireland16a}, and using this we obtain
\begin{align}
\mathcal{P^B}({r},-s\vert{\xi},0)\approx\Bigg\langle \delta\Bigg(\Bigg[{\xi}^2+2\xi\mathcal{G}(-s){w}^p_\parallel(0)+3\Big(\mathcal{G}(-s){w}^p_\parallel(0)\Big)^2\Bigg]^{1/2}-{r}\Bigg)\Bigg\rangle_{{\xi}}.\label{Pr1}
\end{align}
Since the random variable in the argument of $\delta(\cdot)$ in (\ref{Pr1}) is ${w}^p_\parallel(0)$, let us define $\mathcal{F}({w}^p_\parallel(0))\equiv [{\xi}^2+2\mathcal{G}(-s)\xi{w}^p_\parallel(0)+3(\mathcal{G}(-s){w}^p_\parallel(0))^2]^{1/2}-{r}$. Then, using standard results on the properties of $\delta(\cdot)$ we write
\begin{align}
\mathcal{P^B}({r},-s\vert{\xi},0)=\Bigg\langle  \delta\Bigg(\mathcal{F}\Big({w}^p_\parallel(0)\Big) \Bigg)\Bigg\rangle_{{\xi}}=\mathcal{C}\sum_{\mathcal{Z}_n} \Big\vert \mathcal{F}'(\mathcal{Z}_n)\Big\vert^{-1}\Bigg\langle  \delta\Bigg({w}^p_\parallel(0)-\mathcal{Z}_n \Bigg) \Bigg\rangle_{{\xi}},\label{PDFfn}
\end{align}
where  $\mathcal{C}$ is a normalization constant to ensure $\int_0^{\infty}\mathcal{P^B}\,dr=1$, and $\mathcal{Z}_n$ is the $n^{th}$ root of $\mathcal{F}$, i.e. $\mathcal{F}(\mathcal{Z}_n)\equiv 0$, of which there are two
\begin{align}
\mathcal{Z}_1=\frac{-\xi+\sqrt{3r^2-2\xi^2}}{3\mathcal{G}(-s)},\\
\mathcal{Z}_2=\frac{-\xi-\sqrt{3r^2-2\xi^2}}{3\mathcal{G}(-s)}.
\end{align}
The following argument shows that only $\mathcal{Z}_1$ is viable: In the limit $s\to0$, $|2\mathcal{G}(-s)\xi{w}^p_\parallel(0)|\gg |3(\mathcal{G}(-s){w}^p_\parallel(0))^2|$, and solutions with $r>\xi$ must therefore correspond to ${w}^p_\parallel(0)<0$, and solutions with $r<\xi$ must correspond to ${w}^p_\parallel(0)>0$ (recall that $\mathcal{G}(-s)\leq0$). The root $\mathcal{Z}_2$ is not consistent with this; it implies that $r>\xi$ corresponds to ${w}^p_\parallel(0)>0$ . We also note that both roots imply that solutions for $\mathcal{P^B}$ are only real for $r\geq\sqrt{2/3}\xi$. This is because we assumed in (\ref{Pr1}) that the statistics of the longitudinal and perpendicular projections of $\bm{w}^p(0)$ are approximately equal in the caustic regions, and this gave rise to the factor $3$. Let us now suppose that in a given realization they are not equal, and so we would replace $3$ in (\ref{Pr1}) with $\sigma$, where $\sigma\equiv \|\bm{w}^p(0)\|^2/[w^p_\parallel(0)]^2$, such that $\sigma\geq 1$. If we then consider the case $[{\xi}^2+2\xi\mathcal{G}(-s){w}^p_\parallel(0)+\sigma(\mathcal{G}(-s){w}^p_\parallel(0))^2]^{1/2}={r}$, we obtain the solution for ${w}^p_\parallel(0)$
\begin{align}
{w}^p_\parallel(0)=\frac{-\xi\pm\sqrt{\xi^2(1-\sigma)+\sigma r^2}}{\sigma\mathcal{G}(-s)}.
\end{align}
This can be interpreted as the value(s) of ${w}^p_\parallel(0)$ that generate a particle-pair separation equal to $r$ at time $-s$, conditioned on the separation being $\xi$ at time $s=0$. The solution for ${w}^p_\parallel(0)$ is real only if $\sqrt{\xi^2(1-\sigma)+\sigma r^2}\geq0$, i.e. $r\geq \xi\sqrt{(\sigma-1)/\sigma}$. This shows that unless $\sigma=1$, the minimum separation that a particle-pair can reach is $>0$, i.e. the particle trajectories cannot cross. But $\sigma=1$ corresponds to the case where the perpendicular component of the particle relative velocity at $s=0$ is zero. This makes sense; only if the perpendicular component of $\bm{w}^p(0)$ is identically zero can the two-particle trajectories intersect. If the perpendicular component of $\bm{w}^p(0)$ is finite, i.e. $\sigma>1$, then the particle-pair will reach a minimum separation, before they begin to move apart. The implication of this is that to correctly predict $\mathcal{P^B}$ over the range $r\in[0,\xi]$, we would need to account for the range of possible values of $\sigma$ (we will return to this point momentarily when we consider the case of fluid particles). We leave the incorporation of this additional complexity to future work, and focus here on the behavior of separating particle-pairs, i.e. $\mathcal{P^B}$ for $r\geq\xi$.

Following the previous arguments, using only the root $\mathcal{Z}_1$, the result in (\ref{PDFfn}) leads to
\begin{align}
\mathcal{P^B}({r},-s\vert{\xi},0)=\mathcal{C}\vert\mathcal{G}(-s)\vert^{-1}\frac{r}{\sqrt{3r^2-2\xi^2}} \Bigg\langle  \delta\Bigg({w}^p_\parallel(0)-\mathcal{Z}_1 \Bigg) \Bigg\rangle_{{\xi}}.\label{Pr2New}
\end{align}
With this approach, we have translated the problem of predicting $\mathcal{P^B}$ to the problem of predicting $\langle\delta({w}^p_\parallel(0)-\mathcal{Z}_1)\rangle_{\xi}$. However, the latter is simply the PDF of ${w}^p_\parallel(0)$, concerning which a number of asymptotic predictions have been derived. The theoretical work by \citet{gustavsson11,gustavsson14b} predicts the asymptotic result 
\begin{align}
\Big\langle\delta\Big({w}^p_\parallel(0)-w_\parallel\Big)\Big\rangle_{{\xi}}\propto (\xi/\tau_p)^{3-d_2}|w_\parallel|^{d_2-4},\label{GM}
\end{align}
where $d_2(St)$ is the correlation dimension, and $w_\parallel$ is the phase-space variable corresponding to ${w}^p_\parallel(0)$. The result (\ref{GM}) applies when caustics are present, except in the limits $|w_\parallel|\to 0$ and $|w_\parallel|\to \infty$, the latter because the algebraic tail of the PDF is cut off at large $|w_\parallel|$ due to the finitude of the large scale fluid velocity fluctuations. The result in (\ref{GM}) has recently been confirmed using Direct Numerical Simulations \citep{perrin15}, for $St\geq O(1)$, and separations in the dissipation range.

We may use (\ref{GM}) in (\ref{Pr2New}) and obtain 
\begin{align}
\mathcal{P^B}({r},-s\vert{\xi},0)&\propto \Bigg(\frac{\xi\vert\mathcal{G}(-s)\vert}{\tau_p}\Bigg)^{3-d_2} \frac{r}{\sqrt{3r^2-2\xi^2}}\Big\vert\xi- \sqrt{3r^2-2\xi^2}\Big\vert^{d_2-4}.\label{Pr3}
\end{align}
Equation (\ref{Pr3}) predicts that $\mathcal{P^B}$ grows with $s$ as $\vert\mathcal{G}(-s)\vert^{3-d_2}\equiv\vert\tau_p(1-e^{s/\tau_p})\vert^{3-d_2}$, which is a very non-trivial behavior since $d_2(St)$ takes on non-integer values. Indeed, in the limit $s/\tau_p\to0$, the leading order behavior is $\vert\mathcal{G}(-s)\vert^{3-d_2}\sim s^{3-d_2}$. An important point to note, however, is that the extent of $r$-space over which (\ref{Pr3}) will be accurate will itself be a function of time, since (\ref{GM}), upon which (\ref{Pr3}) depends, does not describe the entire distribution of $w^p_\parallel(0)$. The implication of this is that the proportionality constant in (\ref{Pr3}) is then itself implicitly a function of time. Therefore (\ref{Pr3}) can only be used to predict the $r$-dependence, not the $s$ dependence of $\mathcal{P^B}$. However, as a test, we simulated (\ref{Pr1}) for the case where $w^p_\parallel(0)$ is governed entirely by (\ref{GM}), and the results matched (\ref{Pr3}) exactly, confirming that in this regime, the time dependence of $\mathcal{P^B}$ is indeed given by $\vert\mathcal{G}(-s)\vert^{3-d_2}$.

It is straightforward to show that the same analysis for the FIT case leads to the result
\begin{align}
\mathcal{P^F}({r},s\vert{\xi},0)&\propto\Bigg(\frac{\xi\mathcal{G}(s)}{\tau_p}\Bigg)^{3-d_2} \frac{r}{\sqrt{3r^2-2\xi^2}}\Big\vert\xi- \sqrt{3r^2-2\xi^2}\Big\vert^{d_2-4}.\label{Pr3F}
\end{align}
The only difference between (\ref{Pr3}) and (\ref{Pr3F}) is that the former contains $\vert\mathcal{G}(-s)\vert^{3-d_2}$ while the latter contains $\mathcal{G}^{3-d_2}(s)$. These functions satisfy $\vert\mathcal{G}(-s)\vert^{3-d_2}/\mathcal{G}^{3-d_2}(s)\geq1$, and their difference captures the short-time irreversibility produced by the dissipation arising from the drag force on the particles. Consistent with our earlier arguments, these results predict that particles should separate faster BIT than FIT. However, the difference between (\ref{Pr3}) and (\ref{Pr3F}) does not fully account for irreversibility in the short-time regime since we have used (\ref{GM}) which does not capture the asymmetry in the distribution $\langle\delta({w}^p_\parallel(0)-w_\parallel)\rangle_{{\xi}}$. Unfortunately, there is at present no known way to predict the asymmetry of $\langle\delta({w}^p_\parallel(0)-w_\parallel)\rangle_{{\xi}}$.

The results (\ref{Pr3}) and (\ref{Pr3F}) predict that $\mathcal{P^B}$ and  $\mathcal{P^F}$ will change with $Re_\lambda$ through $d_2$. Since it is known that for $St\gtrsim1$, $d_2$ decreases with increasing $Re_\lambda$ (see \citet{ireland16a}, where $c_1\equiv 3-d_2$), then the tails of $\mathcal{P^B}$ and $\mathcal{P^F}$ will decay \emph{faster} with \emph{increasing} $Re_\lambda$. This is the opposite of what we would expect for fluid particles, as we shall soon discuss. It is important to note, however, that the functional forms in (\ref{Pr3}) and (\ref{Pr3F}) are in another sense universal. That is, these functional forms apply even in the absence of any intermittency in the underlying fluid velocity field. This follows from the universality of (\ref{GM}) itself, which holds in both simple Gaussian flow fields \citep{gustavsson11} and also in Navier-Stokes turbulence \citep{perrin15}. These results therefore suggest that extreme events (here characterized by large separations over short timescales) in the pair-separation of inertial particles with $St\geq O(1)$ occur, not as a consequence of the intermittency in the small-scale turbulence itself, but as a consequence of their strongly non-local dynamics. 

Next, we consider $St=0$ and arbitrary $\xi$. In this case, following \citet{batchelor52a}, we may use a short-time expansion in $s$ and obtain
\begin{align}
\bm{r}^p(-s)&\approx\bm{r}^p(0)-s\Delta\bm{u}^p(0)+O(s^2),\label{rfsol}
\end{align}
where $\Delta\bm{u}^p(0)\equiv\Delta\bm{u}(\bm{r}^p(0),0)$, leading to
\begin{align}
\mathcal{P^B}({r},-s\vert{\xi},0)\approx\Big\langle \delta\Big(\|\bm{\xi}-s\Delta\bm{u}^p(0)\|-{r}\Big)\Big\rangle_{{\xi}}.\label{Pb1}
\end{align}
A significant complication compared to the $St\geq O(1)$ case is that the parallel and perpendicular components of $\Delta\bm{u}^p(0)$ are not statistically equivalent in general, or even approximately so. Therefore, if we write 
\begin{align}
\|\Delta\bm{u}^p(0)\|^2=\mu^p[\Delta u^p_\parallel(0)]^2,\label{upa}
\end{align}
then the coefficient $\mu^p\geq1$ will be random, varying significantly from one realization of the ensemble to another. In principle, we could capture the effect of $\mu^p$ on $\mathcal{P^B}$ in the short-time limit by using (\ref{upa}) in (\ref{Pb1}) to obtain

\begin{align}
\mathcal{P^B}({r},-s\vert{\xi},0)\approx\Big\langle \delta\Big(\Big[\xi^2-2s\xi\Delta{u}^p_\parallel(0)+\mu^p\Big(s\Delta u^p_\parallel(0)\Big)^2\Big]^{1/2} -{r}\Big)\Big\rangle_{{\xi}},\label{pdf1}
\end{align}
and then apply to this the decomposition
\begin{align}
\mathcal{P^B}({r},-s\vert{\xi},0)\equiv\int_\mu \mathcal{P^B}({r},-s\vert{\xi},0,\mu)\varphi(\mu)\,d\mu,\label{PBdecomp}
\end{align}
where
\begin{align}
\mathcal{P^B}({r},-s\vert{\xi},0,\mu)&=\Big\langle \delta\Big(\Big[\xi^2-2s\xi\Delta{u}^p_\parallel(0)+\mu^p\Big(s\Delta u^p_\parallel(0)\Big)^2\Big]^{1/2} -{r}\Big)\Big\rangle_{{\xi},\mu^p=\mu}\label{Pb2}\\
\varphi(\mu)&\equiv \Big\langle\delta\Big(\mu^p-\mu\Big)\Big\rangle,
\end{align}
$\mu$ being the sample-space variable corresponding to $\mu^p$. In general we cannot evaluate (\ref{PBdecomp}) since we do not know $\varphi(\mu)$. However, from (\ref{rfsol}) we find that to leading order in $s$, $\|\bm{r}(-s)\|^2-\xi^2\propto s\Delta u_\parallel^p(0)$, and we might expect that $\Delta u_\parallel^p(0)$ is typically larger in regions where $\mu=O(1)$ than in regions where $\mu\gg1$. For example in regions of strong, quasi-solid body rotation where $\mu\gg 1$, $\Delta u_\parallel^p(0)$ would be very small compared with its typical values in regions dominated by extensional straining where $\mu=O(1)$. This then implies that $\mathcal{P^B}$ should be dominated by the contributions with $\mu=O(1)$ in the limit $s\to0$. As an approximation we assume that the statistics of the parallel and perpendicular components of $\Delta\bm{u}^p(0)$ are equal, corresponding to assuming $\varphi(\mu)=\delta(3-\mu)$ in (\ref{PBdecomp}). Then, following the same procedure as was used to derive the result for the $St\geq O(1)$ case, we obtain
\begin{align}
\mathcal{P^B}({r},-s\vert{\xi},0)=\mathcal{C}\frac{r}{s\sqrt{3r^2-2\xi^2}} \Bigg\langle  \delta\Bigg(\Delta{u}^p_\parallel(0)- \mathcal{Z}  \Bigg) \Bigg\rangle_{{\xi}},\label{Pr2Newb}
\end{align}
where
\begin{align}
\mathcal{Z}=\frac{\xi-\sqrt{3r^2-2\xi^2}}{3 s}.
\end{align}
Since we are considering a statistically stationary system where the fluid particles are fully-mixed (globally), then $\langle\delta(\Delta{u}^p_\parallel(0)- \mathcal{Z}) \rangle_{{\xi}}$ is identical to the PDF of the fluid velocity differences at scale $\xi$, i.e. $\langle\delta(\Delta{u}^p_\parallel(0)- \mathcal{Z}) \rangle_{{\xi}}=\langle\delta(\Delta{u}_\parallel(\xi,0)- \mathcal{Z}) \rangle$, and this can be described using the multifractal formalism. In order to capture the asymmetry of this PDF, we use the multifractal model from \citet{chevillard12}
\begin{align}
\begin{split}
\Bigg\langle  \delta\Bigg(\Delta{u}_\parallel(\xi,0)- \mathcal{Z} \Bigg) \Bigg\rangle_{{\xi}}=\int_{h_{min}}^{h_{max}}\frac{1}{\sigma\beta_\xi}\mathcal{P}_{h,\xi}\mathcal{P}_\delta\Bigg( \frac{\mathcal{Z}}{\sigma\beta_\xi} \Bigg)\, dh,\label{multifrac}
\end{split}
\end{align}
where 
\begin{align}
\beta_\xi(h)&\equiv (\xi/L)^h \Big(1+(\eta_h/\xi)^{2}\Big)^{(h-1)/2},\\
\mathcal{P}_{h,\xi}(h)&\equiv A^{-1}  (\xi/L)^{1-\mathcal{D}_h}  \Big(1+(\eta_h/\xi)^{2}\Big)^{(1-\mathcal{D}_h)/2},\\
\mathcal{P}_\delta(y)&\equiv\frac{1}{\sqrt{2\pi}}\Bigg(e^{-y^2/2}-\lambda_\xi y(y^2-1)e^{-y^2/2a^2}\Bigg),
\end{align}
${A}$ is a normalization constant, ${h_{min}=0}$, ${h_{max}=1}$, ${\mathcal{D}_h=1-(h-c_1)^2/2 c_2}$, ${a=\sqrt{9/10}}$, ${c_2=1/40}$, ${c_1=(1/3)+(3 c_2/2)}$, $\sigma=\sqrt{2}u'$, ${\eta_h=L(Re/R^{*})^{-1/(h+1)}}$, $L$ is the integral lengthscale, $Re=\sigma L/\nu$, and $R^*=52$. The parameter $\lambda_\xi$ allows the model to capture the asymmetry of  $\langle\delta(\Delta{u}_\parallel(\xi,0)- \mathcal{Z}) \rangle$, and is given by $\lambda_\xi=(2\langle\epsilon\rangle \xi/15-\nu\nabla_\xi\langle[\Delta u_\parallel(\xi,0)]^2\rangle)/\langle\beta_\xi^3\rangle$, where $\langle\beta_\xi^3\rangle\equiv\int_{h_{min}}^{h_{max}}\beta^3_\xi(h)\mathcal{P}_{h,\xi}(h)\,dh$.

In the multifractal theory, $h$ plays the role of the scaling exponent, describing how $\Delta u_\parallel(\xi,0)$ scales with $\xi$ on a given fractal subset of the system. The exponent $h$ has the PDF $\mathcal{P}_{h,\xi}(h)$, and the associated singularity spectrum is $\mathcal{D}_h$. The result in  (\ref{multifrac}) captures the transition from an essentially Gaussian PDF for $\Delta u_\parallel(\xi,0)$ at $\xi=L$, to a highly non-Gaussian PDF at $\xi=\eta$, exhibiting heavy, stretched exponential-type decaying tails. It is important to note that \eqref{multifrac} is based upon the ``log-normal'' assumption \citep{kolmogorov62}, which is known to lead to unphysical descriptions of the most extreme fluctuations of the fluid velocity field \citep[see][]{frisch}. As a result, using \eqref{multifrac} in \eqref{Pr2Newb} could lead to a failure to correctly describe the dispersion of particle-pairs that separate extremely fast. Nevertheless, the use of the multifractal model of \citet{chevillard12} is just a choice; if future multifractal models for $\langle\delta(\Delta{u}_\parallel(\xi,0)- \mathcal{Z}) \rangle$ emerge that capture both the asymmetry of the PDF and go beyond the log-normal assumption, they could be used in \eqref{Pr2Newb} instead. We refer the reader to \citet{benzi91,frisch,boffetta08} for detailed discussions of the multifractal formalism in turbulence.

Using (\ref{multifrac}) in (\ref{Pr2Newb}) we obtain 
\begin{align}
\mathcal{P^B}({r},-s\vert{\xi},0)=\int_{h_{min}}^{h_{max}}\frac{\mathcal{C}r\mathcal{P}_{h,\xi}}{\sigma\beta_\xi s\sqrt{3r^2-2\xi^2}}\mathcal{P}_\delta\Bigg( \frac{\xi-\sqrt{3r^2-2\xi^2}}{3\sigma\beta_\xi s} \Bigg)\, dh
.\label{Pr2Newb2}
\end{align}
As expected, the result in (\ref{Pr2Newb2}) shows that for $r=O(\xi)$, the shape of $\mathcal{P^B}$ is affected by the initial condition, but for $r\gg\xi$, the shape of $\mathcal{P^B}$ reflects the shape of the underlying PDF for $\Delta u_\parallel(\xi,t)$. It therefore also describes the variation in the behavior of $\mathcal{P^B}$ in the short-time regime as $\xi$ is varied, transitioning from an approximately Gaussian PDF for $\xi=O(L)$ to a highly non-Gaussian PDF for $\xi\leq O(\eta)$.

The corresponding prediction for $\mathcal{P^F}$ is
\begin{align}
\mathcal{P^F}({r},s\vert{\xi},0)=\int_{h_{min}}^{h_{max}}\frac{\mathcal{C}r\mathcal{P}_{h,\xi}}{\sigma\beta_\xi s\sqrt{3r^2-2\xi^2}}\mathcal{P}_\delta\Bigg( \frac{-\xi+\sqrt{3r^2-2\xi^2}}{3\sigma\beta_\xi s} \Bigg)\, dh
.\label{PFpred}
\end{align}
Since $\mathcal{P}_\delta$ is asymmetric, it follows that  (\ref{Pr2Newb2}) and (\ref{PFpred}) predict $\mathcal{P^B}\neq\mathcal{P^F}$, consistent with our earlier arguments that the irreversibility of fluid particle-pair dispersion arises because the PDF of $\Delta u_\parallel$ is asymmetric. More specifically, $\mathcal{P}_\delta$ gives rise to a negatively skewed PDF for $\Delta u_\parallel$, such that (\ref{Pr2Newb2}) and (\ref{PFpred}) predict the fluid particles separate faster BIT than FIT, in agreement with the arguments of \S\ref{IM}.

The multifractal prediction for the PDF of $\Delta u_\parallel$ captures the enhanced intermittency as $Re_\lambda$ is increased. Through (\ref{Pr2Newb2}) this leads to the prediction that the tails of $\mathcal{P^B}$ decay more slowly as $Re_\lambda$ is increased, as expected. What is interesting is that this is in contrast to the behavior of $St\gtrsim 1$ particles, for which, according to (\ref{Pr3}), the tails of $\mathcal{P^B}$ decay \emph{faster} as $Re_\lambda$ is increased.

We now consider the regime $s/\tau^p_\xi\gg1$ (but finite), in which most of the particles will be found in the inertial range of the turbulence if $Re_\lambda\to\infty$. For fluid particles in this regime, a famous prediction of Richardson is that $\mathcal{P^F}$ should be independent of $\xi$ with the distribution (see \citet{salazar09})
\begin{align}
\mathcal{P^F}({r},s\vert{\xi},0)\propto \frac{r^2}{\langle\|\bm{r}^p(s)\|^2\rangle_{\xi}^{3/2}}\exp\Bigg[ -\frac{\mathcal{A^F} r^{2/3}}{\langle\|\bm{r}^p(s)\|^2\rangle_{\xi}^{1/3}}\Bigg],\label{Rpdf}
\end{align}
where $\langle\|\bm{r}^p(s)\|^2\rangle_{\xi}\propto \langle\epsilon\rangle s^3$, and $\mathcal{A^F}>0$ is a constant. The most recent tests of the accuracy of (\ref{Rpdf}) have demonstrated that it describes $\mathcal{P^F}$ quite well for fluid particles at long times when the particles are in the inertial range \citep{bitane13,biferale14}, except for very small and very large $r$, which correspond to extreme events in the pair-separation process  \citep{scatamacchia12,biferale14}. Under the phenomenology employed by Richardson we also have
\begin{align}
\mathcal{P^B}({r},-s\vert{\xi},0)\propto \frac{r^2}{\langle\|\bm{r}^p(-s)\|^2\rangle_{\xi}^{3/2}}\exp\Bigg[ -\frac{\mathcal{A^B} r^{2/3}}{\langle\|\bm{r}^p(-s)\|^2\rangle_{\xi}^{1/3}}\Bigg],\label{RpdfBIT}
\end{align}
where $\langle\|\bm{r}^p(-s)\|^2\rangle_{\xi}\propto \langle\epsilon\rangle s^3$, and $\mathcal{A^B}>0$ is a constant. 

The form of Richardson's PDF follows from the assumption that the pair-separation is governed by a diffusion process, with a diffusion coefficient
\begin{align}
\mathcal{D}(r)\propto r^{4/3},
\end{align}
which can also be arrived at using K41 scaling. Let us now consider how the shape of the Richardson PDF might be modified for inertial particles through the effects of the inertia on the scaling of $\mathcal{D}(r)$. In the regime $St_r\ll1$, it is straightforward to show that $\mathcal{D}(r)\propto r^{4/3}+O[St_r]$ \citep{bragg14e}. For $St_r\geq O(1)$, it is more difficult to determine the scaling of $\mathcal{D}$ since the inertial particle relative velocity statistics in the inertial range do not in general scale with $r$ as simple power laws. Results from PDF theories for inertial particles predict that \citep{zaichik09,bragg14b}
\begin{align}
\mathcal{D}(r)=\tau_p\lambda+\tau_p\Big\langle w^p_\parallel(t) w^p_\parallel(t)\Big\rangle_{r},\label{ZTD}
\end{align}
where 
\begin{align}
\lambda(r)=\frac{1}{St_r(1+St_r)}\Big\langle\Delta{u}_\parallel(r,t)\Delta{u}_\parallel(r,t)\Big\rangle,
\end{align}
is a diffusion coefficient that describes the effect of $\Delta u_\parallel(r,t)$ on the transport of particles in $r$-space \citep{bragg14b}. When $St_r\ll1$, $\lambda\propto r^{4/3}$, but when $St_r\gg1$ $\lambda\propto r^{2}$.

In the inertial range, the effect of the non-local inertial particle dynamics on $\langle w^p_\parallel w^p_\parallel\rangle_{r}$ is quite weak when $St_r\leq O(1)$ compared with the effect of inertial filtering \citep{ireland16a}. We may then approximate $\langle w^p_\parallel w^p_\parallel\rangle_{r}$ based on its local behavior \citep{zaichik09,bragg14c}
\begin{align}
\Big\langle w^p_\parallel(t) w^p_\parallel(t)\Big\rangle_{r}\approx\frac{1}{1+St_r}\Big\langle\Delta{u}_\parallel(r,t)\Delta{u}_\parallel(r,t)\Big\rangle.\label{w2local}
\end{align}
Provided then that $St_r\leq O(1)$ in the inertial range, substituting (\ref{w2local}) in (\ref{ZTD}) gives
\begin{align}
\mathcal{D}(r)\approx\tau_r\Big\langle\Delta{u}_\parallel(r,t)\Delta{u}_\parallel(r,t)\Big\rangle,
\end{align}
which becomes $\mathcal{D}(r)\propto r^{4/3}$ under K41 scaling. These considerations therefore suggest that in the regime for which Richardson's PDF shape holds true for fluid particles, the same PDF shape should also apply even up to $St_r=O(1)$. Of course we have neglected to consider here the fact that Richardson's Markovian assumption will in principle be even less applicable for finite $St_r$ than it is for fluid particles. We will not consider this here, but it is an issue that we shall address in a future publication.

Finally, in the limit $s/\tau^p_\xi\to\infty$, we expect the state of the system to be independent of its state at $s=0$, giving the time-independent distributions\footnote[1]{In (\ref{Bgrscal}) we have assumed that the single time probability measures of the system are time invariant on the time interval $(-\infty,+\infty)$. This is not necessarily true since any real dynamical system will pass through an initial transient regime. However, in reality we expect the asymptotic scaling (\ref{Bgrscal}) to be realized for $-s\ll-\tau_I$, where $\tau_I$ is the integral timescale for the system. Therefore, if we denote the initial time of the system by $-T_0$ (e.g. the time at which the experiment began), then provided $-T_0\lll-\tau_I$, the result in (\ref{Bgrscal}) makes sense for $-T_0\ll-s\ll-\tau_I$.}
\begin{align}
\lim_{s\to\infty}\mathcal{P^F}({r},s\vert{\xi},0)&\propto r^2g^{\infty}(r),\label{Fgrscal}\\
\lim_{s\to\infty}\mathcal{P^B}({r},-s\vert{\xi},0)&\propto r^2g^{\infty}(r),\label{Bgrscal}
\end{align}
where $g^{\infty}(r)$ is the stationary form of the Radial Distribution Function (RDF). Since $g^{\infty}(r)$ is strongly dependent upon $St$, the asymptotic state of the particle-pairs will vary with $St$. Furthermore, for $St>0$, the asymptotic forms of $\mathcal{P^F}$ and $\mathcal{P^B}$ will in principle depend upon $Re_\lambda$ through $g^{\infty}(r)$, e.g. see \citet{ireland16a}.
\section{DNS Results \& Discussion}\label{RaD}
\begin{figure}
\centering
\vspace{-30mm}
{\begin{overpic}
[trim = 15mm 100mm 15mm 20mm,scale=0.8,clip,tics=20]{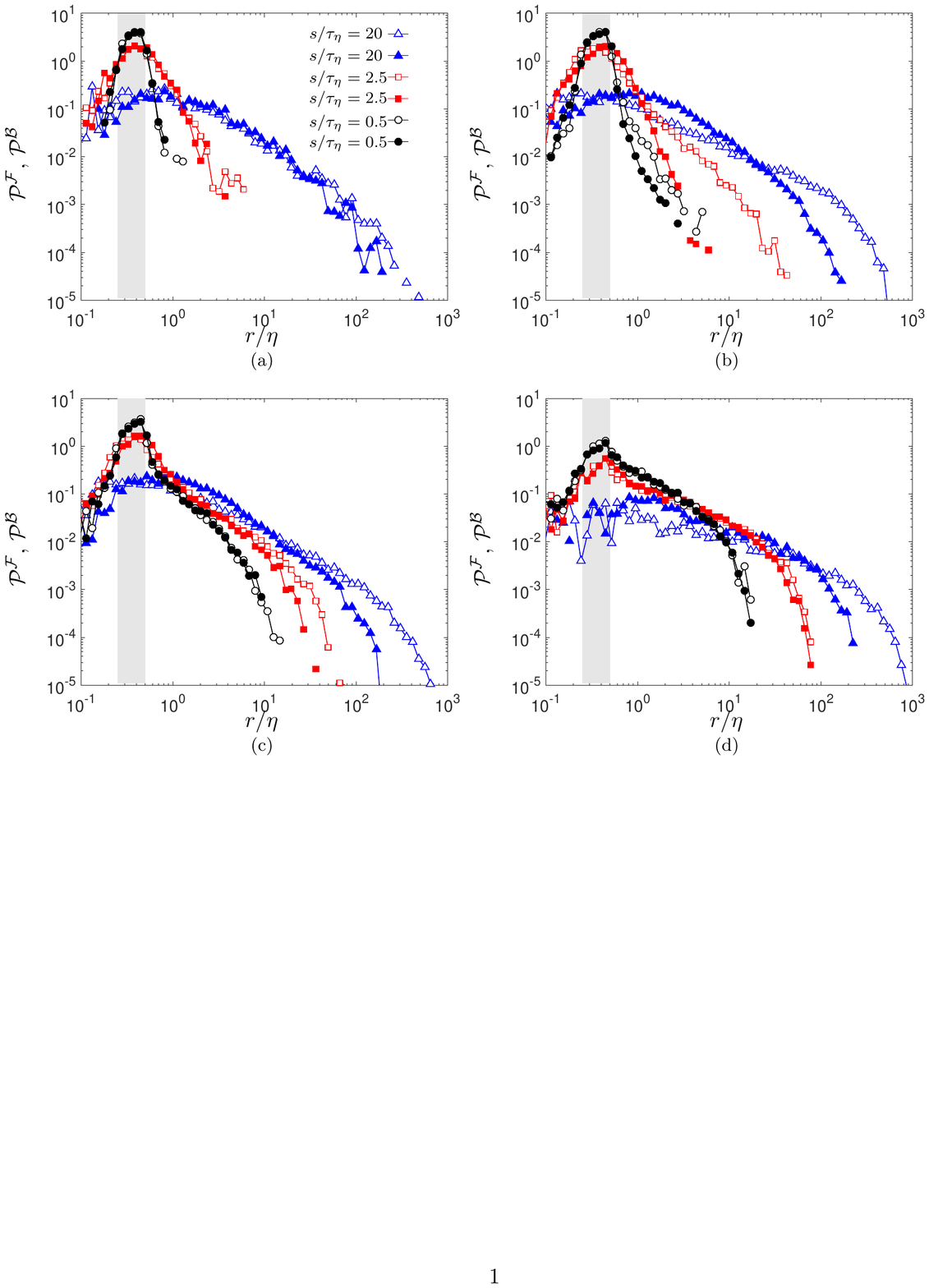}
\end{overpic}}
\caption{DNS results for $\mathcal{P^F}$ (filled symbols) and $\mathcal{P^B}$ (open symbols)  with $\xi/\eta\in[0.25,0.5]$ and (a) $St=0$, (b) $St=0.5$, (c) $St=3$, (d) $St=10$. The grey box indicates the set of values $\xi/\eta\in[0.25,0.5]$, corresponding to the initial/final separation of the particle-pair.}
\label{Ssep1}
\end{figure}
\FloatBarrier
We now come to test the theoretical arguments and results from \S\ref{theory}, and to further explore the problem using results from Direct Numerical Simulations (DNS) of particle-laden, statistically stationary, homogeneous, isotropic turbulence. The DNS dataset is identical to that in \citet{ireland16a}, and we therefore refer the reader to that paper for the details of the DNS; here we summarize. A pseudospectral method was used to solve the incompressible Navier-Stokes equations on a periodic domain of length $2\pi$, with $2048^3$ grid points, using $16384$ processors on the Yellowstone cluster at the U.S. National Center for Atmospheric Research \citep{yellowstone}. Deterministic forcing was applied in Fourier space at low wavenumbers to generate statistically stationary, homogeneous, isotropic turbulence. The Taylor Reynolds number $Re_\lambda$ for our flow is about $582$, and $L/\eta \approx 800$, where $L$ is the integral lengthscale. Inertial particles subject to a linear drag force (corresponding to that used in the theory in \S\ref{theory}) were then tracked in this flow field. A total of 18 different particle classes were simulated, with $St\in[ 0,30]$, with about $17 \times 10^6$ particles tracked for each value of $St$. After the flow field had become statistically stationary, the particles were injected in the flow with a uniform distribution. The particle dispersion statistics were only computed after their RDFs and velocities had reached a statistically stationary state. 

\begin{figure}
\centering
\vspace{-7mm}
{\begin{overpic}
[trim = 15mm 100mm 15mm 20mm,scale=0.8,clip,tics=20]{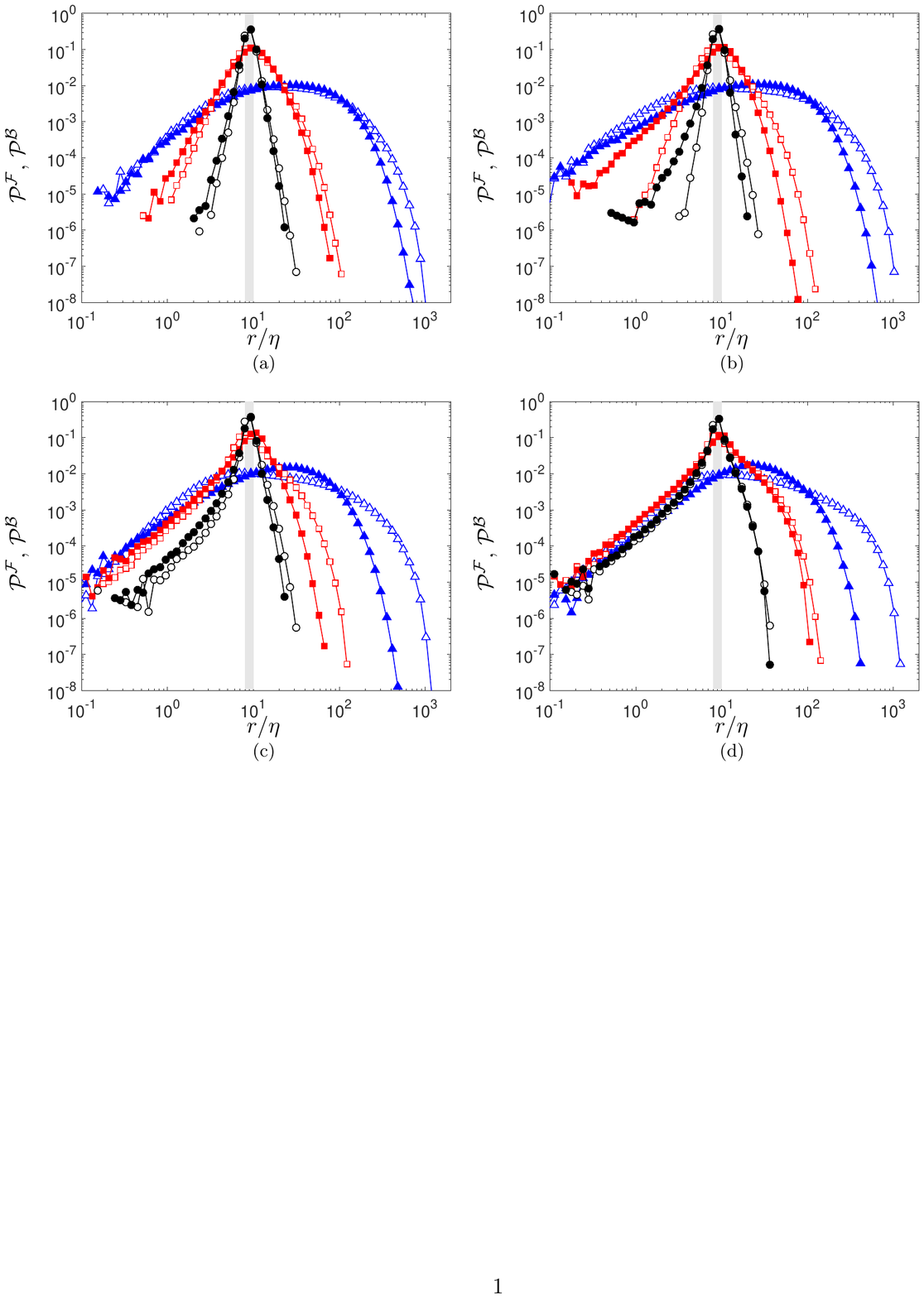}
\end{overpic}}
\caption{DNS results for $\mathcal{P^F}$ (filled symbols) and $\mathcal{P^B}$ (open symbols)  with $\xi/\eta\in[8,10]$ and (a) $St=0$, (b) $St=0.5$, (c) $St=3$, (d) $St=10$. Legend is the same as that in Figure~\ref{Ssep1}.}
\label{Lsep1}
\end{figure}

We begin by considering the dispersion irreversibility in figure~\ref{Ssep1}, where we show the DNS results for $\mathcal{P^B},\,\mathcal{P^F}$ for particles with initial separation $\xi/\eta\in[0.25,0.5]$. In agreement with the arguments in 
\S\ref{theory}, the results show that the particles separate faster BIT than FIT. The results also show that the irreversibility is strongest for $St=O(1)$, just as was found in \citet{bragg16} when we analyzed the mean-square separation FIT and BIT. Figure~\ref{Lsep1} shows the corresponding results for $\xi/\eta\in[8,10]$. These results confirm the arguments of \S\ref{theory}, that $\mathcal{P^B}>\mathcal{P^F}$ for $r>\xi$, but $\mathcal{P^B}<\mathcal{P^F}$ for $r<\xi$. The confirmation of this non-trivial prediction lends strong support for the validity of the irreversibility mechanisms we have set forth, being able to predict not only the existence of irreversibility in the dispersion, but also its nature, namely whether FIT or BIT dispersion will be fastest. This result also demonstrates that in the short-time limit, the irreversibility is not simply generated by dissipation but also by the asymmetry in the PDF of ${w}_\parallel^p(0\vert\xi,0)$. This is clear since dissipation of kinetic energy in the absence of asymmetry in the PDF of ${w}_\parallel^p(0\vert\xi,0)$ would always trivially cause particles to disperse faster BIT than FIT, irrespective of whether $r>\xi$ or  $r<\xi$.

\begin{figure}
\centering
\vspace{-0mm}
{\begin{overpic}
[trim = 15mm 140mm 15mm 0mm,scale=0.8,clip,tics=20]{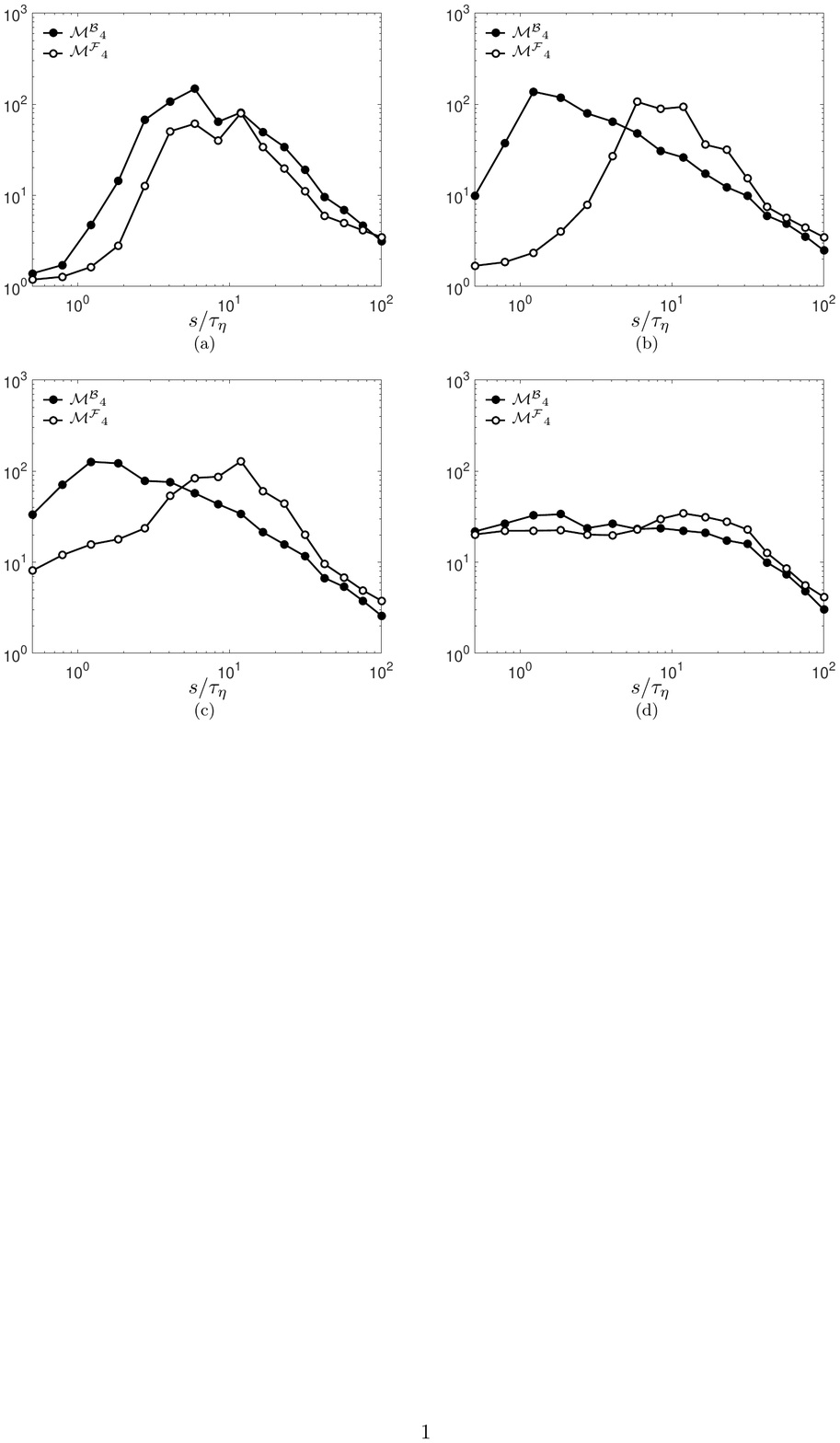}
\end{overpic}}
	\caption{DNS results for $\mathcal{M^F}_4(\xi,s)$ and $\mathcal{M^B}_4(\xi,s)$ with (a) $St=0$, (b) $St=0.5$, (c) $St=1$, (d) $St=3$ and $\xi/\eta\in[0.25,0.5]$.}
\label{M4}
\end{figure}

\begin{figure}
	\centering
	\vspace{-0mm}
	{\begin{overpic}
			[trim = 15mm 140mm 15mm 0mm,scale=0.8,clip,tics=20]{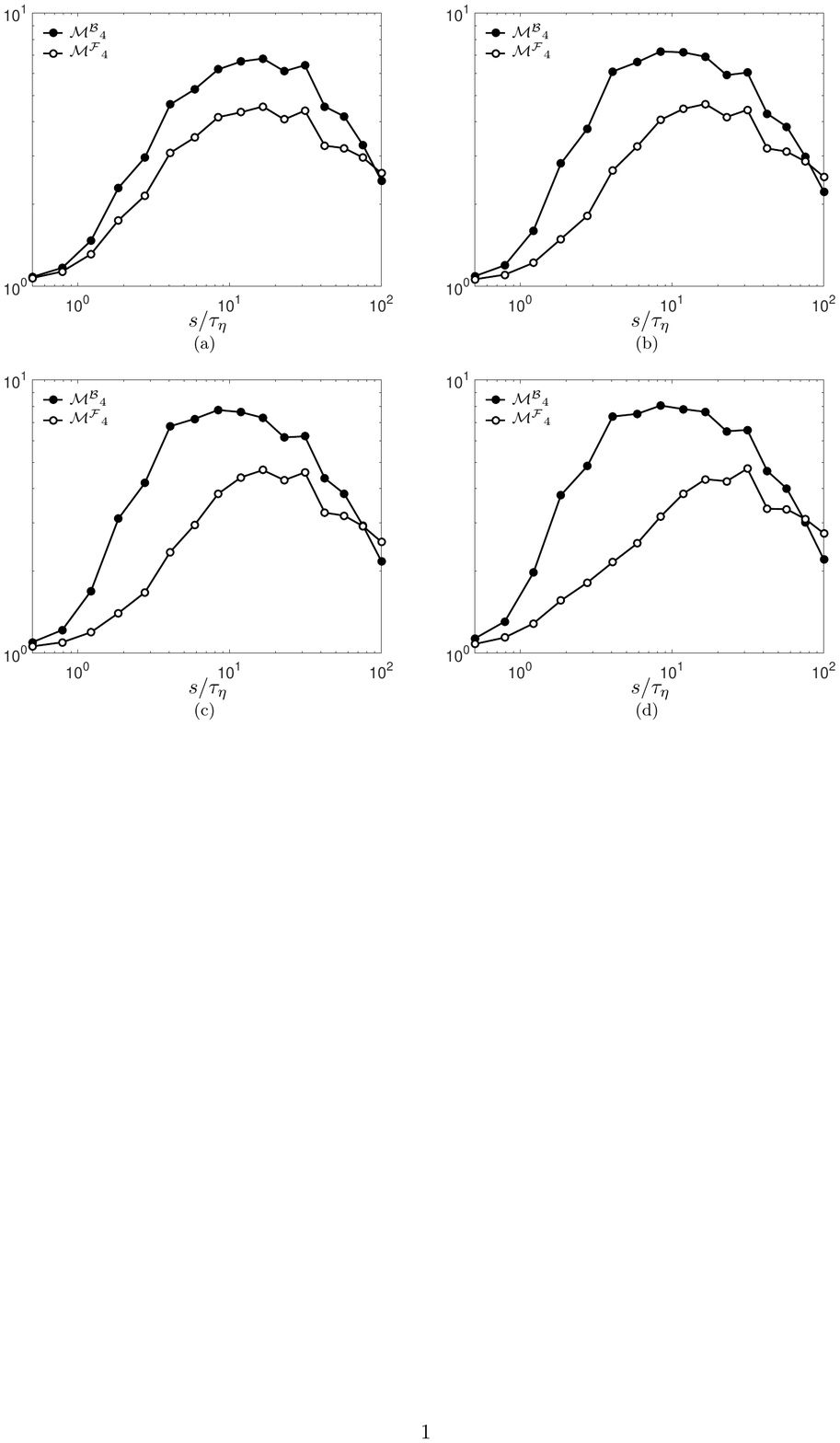}
	\end{overpic}}
	\caption{DNS results for $\mathcal{M^F}_4(\xi,s)$ and $\mathcal{M^B}_4(\xi,s)$ with (a) $St=0$, (b) $St=0.5$, (c) $St=1$, (d) $St=3$ and $\xi/\eta\in[8,10]$.}
	\label{M4b}
\end{figure}

The overall picture of the results in figure~\ref{Ssep1} and figure~\ref{Lsep1} is that FIT, the overwhelming probability is that particles that have the initial separation $\xi$ will go to larger separations, and BIT, the overwhelming probability is that particles that have the final separation $\xi$ came from larger separations. However, their approach to these larger separations is much faster BIT than FIT. One practical consequence of this is that care must be taken in modeling correctly the nature of the particle dispersion in a give problem of interest. For example, the statistics of the evolution of scalar concentration fields in turbulence can be described through integrals of the concentration field measured BIT along particle trajectories \citep{buaria16}. A model that assumed BIT=FIT would lead to significant underpredictions of the mixing of the concentration. Similarly, it has recently been shown that a correct handling of the dependence of inertial particle relative velocities on the BIT dispersion of the particles \citep{pan10} is crucial in order to accurately model the collision velocities of inertial particles in turbulence \citep{bragg17}.
\FloatBarrier
In order to more fully quantify the irreversibility, we can consider the normalized moments of the PDFs
	\begin{align}
	\mathcal{M^F}_N(\xi,s)&\equiv \int_0^\infty r^N\mathcal{P^F}\,dr\Bigg/ \Bigg(\int_0^\infty r^2\mathcal{P^F}\,dr\Bigg)^{N/2},\label{NMF}\\
	\mathcal{M^B}_N(\xi,s)&\equiv \int_0^\infty r^N\mathcal{P^B}\,dr\Bigg/\Bigg(\int_0^\infty r^2\mathcal{P^B}\,dr\Bigg)^{N/2}.\label{NMB}
	\end{align}
	In \citet{bragg16} we considered the ratio of the variances of the BIT to FIT PDFs and showed that for $\xi$ in the dissipation range the ratio could reach values up to $O(10^2)$, showing that BIT dispersion can be much faster than FIT dispersion. Unlike comparing the FIT and BIT variances, comparing (\ref{NMF}) and (\ref{NMB}) does not necessarily give a measure of the difference in the rate at which the particles disperse FIT and BIT because of the normalization. Indeed there is some ambiguity in how to appropriately quantify ``the degree of irreversibility'' associated with the higher order moments of the PDFs. However, comparing (\ref{NMF}) and (\ref{NMB}) can provide insight concerning how the ``shape'' of the dispersion PDFs differ FIT and BIT. In Figures~\ref{M4} and~\ref{M4b} we plot $\mathcal{M^F}_4(\xi,s)$ and $\mathcal{M^B}_4(\xi,s)$ at different values of $\xi$ and for different $St$ numbers. We note that by definition,  $\mathcal{M^F}_N(\xi,0)\equiv\mathcal{M^B}_N(\xi,0)\equiv1$, and for Gaussian PDFs, $\mathcal{M^F}_4(\xi,s)\equiv\mathcal{M^B}_4(\xi,s)\equiv 3$.
\FloatBarrier
\FloatBarrier
In agreement with the results in \citet{biferale14}, the results in Fig.~\ref{M4} show that for fluid particles with $\xi<\eta$,  $\mathcal{M^F}_4$ peaks at $O(10^2)$ when $s/\tau_\eta=O(10)$. At longer times, $\mathcal{M^F}_4$ decays, associated with the fact that as $s$ increases, most of the particles go to larger separations where the non-Gaussianity of velocity difference PDFs decreases. In an unbounded homogeneous turbulent flow we would expect $\mathcal{M^F}_4\approx 3$ for $s/\tau_\eta\to\infty$, since the large-scale fluid velocity field has an approximately Gaussian PDF. The results in Fig.~\ref{M4} and Fig.~\ref{M4b} also show that for $St=0$, $\mathcal{M^F}_4<\mathcal{M^B}_4(\xi,s)$, implying that the BIT PDF is more intermittent than the FIT PDF. Related to our earlier arguments, this follows because the negative values of $\Delta u_\parallel$ have more extreme behavior than the positive values. For the inertial particles, the results show that inertia affects both the peak values of $\mathcal{M^F}_4$ and $\mathcal{M^B}_4$, and also the time it takes for these quantities to reach their peak values. Figure~\ref{M4}, for example, shows that for $St=1$,  $\mathcal{M^F}_4$ and $\mathcal{M^B}_4$ grow considerably faster than they do for $St=0$. This is associated with the very rapid separation of the inertial particles that arises because of the caustics in their velocity distributions. The results also show that inertia can considerably enhance the asymmetry between the FIT and BIT dispersion PDFs, with inertia leading to considerable differences in the intermittency of the FIT and BIT pair-dispersion. Furthermore, comparing Fig.~\ref{M4} and Fig.~\ref{M4b} shows that, as expected, the non-Gaussianity of the dispersion PDFs decreases as $\xi$ is increased.

Unfortunately, the statistical convergence of our DNS data is not sufficient to consider $\mathcal{M^F}_N(\xi,s)$ and $\mathcal{M^B}_N(\xi,s)$ for $N>4$. However, in a future study we will perform DNS with much larger numbers of particles, that will allow us to probe more fully the role of irreversibility in the most extreme events in the particle-pair dispersion, along with an investigation into the role of $Re_\lambda$.

\begin{figure}
\centering
{\begin{overpic}
[trim = 15mm 200mm 15mm 0mm,scale=0.8,clip,tics=20]{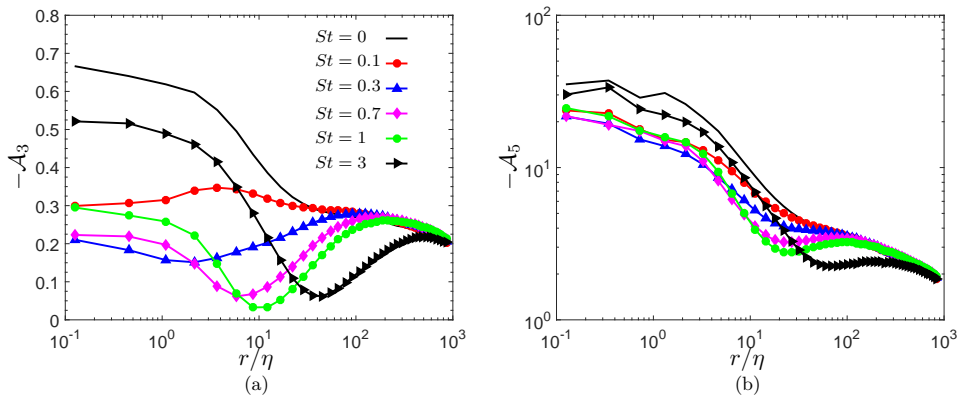}
\end{overpic}}
\caption{DNS results for (a) $\mathcal{A}_3$ and (b) $\mathcal{A}_5$ for various $St$ as a function of $r/\eta$.}
\label{PrefSamp}
\end{figure}

\begin{figure}
\centering
{\begin{overpic}
[trim = 15mm 200mm 15mm 0mm,scale=0.8,clip,tics=20]{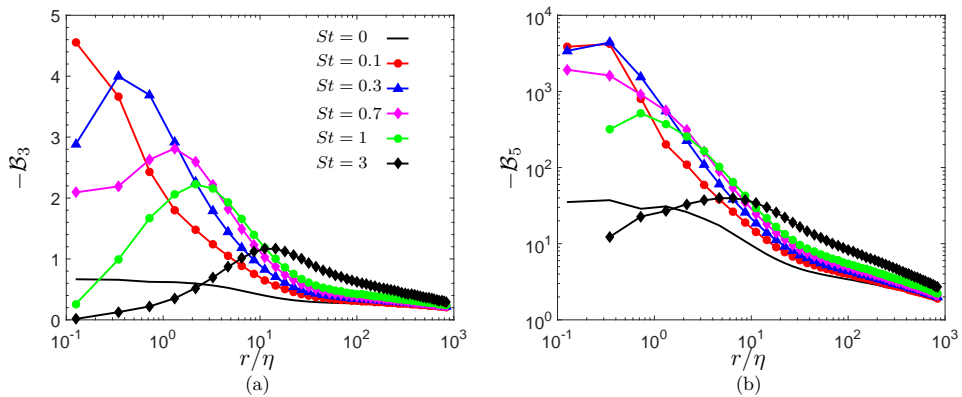}
\end{overpic}}
\caption{DNS results for (a) $\mathcal{B}_3$ and (b) $\mathcal{B}_5$ for various $St$ as a function of $r/\eta$.}
\label{Sp3Sp5}
\end{figure}

As explained in \S\ref{IM}, one of the assumptions made in our exposition of the local and non-local irreversibility mechanisms is that they are affected by preferential sampling only quantitatively, not qualitatively. In particular, we assumed that the properties of $\Delta\bm{u}$ measured along inertial particle trajectories are quantitatively different from those along fluid particle trajectories, but not qualitatively different. To test this assumption, we can consider the normalized moments
\begin{align}
\mathcal{A}_N(r,t)\equiv \Big\langle\Big[\Delta u_\parallel (r^p(t),t)\Big]^N\Big\rangle_r\Bigg/ \Big\langle\Big[\Delta u_\parallel (r^p(t),t)\Big]^2\Big\rangle_r^{N/2}.\label{AN}
\end{align}
These can be used to quantify the role of preferential sampling on) the irreversibility mechanisms by considering its variation with $St$ for odd $N$. In Figure~\ref{PrefSamp} we plot $\mathcal{A}_3$ and $\mathcal{A}_5$, that is, the skewness and hyperskewness of the fluid velocity differences sampled by the inertial particles. The results confirm the expectation; though both $\mathcal{A}_3$ and $\mathcal{A}_5$ are strongly affected by $St$, they remain qualitatively similar in the sense that their sign remains negative. Furthermore, we checked that the non-normalized moments $\langle[\Delta u_\parallel (r^p(t),t)]^N\rangle_r$ all increase monotonically with increasing $r$. Therefore, preferential sampling does not change the nature of the irreversibility; BIT separation remains faster than FIT separation. A most important observation, however, is that preferential sampling \emph{reduces} the asymmetry of the PDF of $\Delta u_\parallel (r^p(t),t)$. This means that the enhanced irreversibility observed for inertial particle-pair dispersion does not arise from preferential sampling. 
\FloatBarrier
This point may be further emphasized by considering the normalized moments of $w^p_\parallel$
\begin{align}
\mathcal{B}_N(r,t)\equiv \Big\langle\Big[w^p_\parallel(t)\Big]^N\Big\rangle_r\Bigg/ \Big\langle\Big[w^p_\parallel(t) \Big]^2\Big\rangle_r^{N/2}.\label{BN}
\end{align}
The results in Figure~\ref{Sp3Sp5} show that inertia can lead to a dramatic enhancement of $\mathcal{B}_3$ and $\mathcal{B}_5$, as compared to the fluid particle case. That is, the skewness and hyperskewness of the PDF of $w^p_\parallel$ can be much larger (in magnitude) than those of the underlying field. When compared with the results in Figure~\ref{PrefSamp}, it becomes evident that the strong increase in $\mathcal{B}_3$ and $\mathcal{B}_5$, does not arise from preferential sampling, but from the non-local contribution to the inertial particle dynamics. Taken together with our earlier arguments, these results support our physical picture, that the enhanced irreversibility of inertial particle-pair dispersion arises from the NLIM, with a contribution at short-times from dissipation induced irreversibility, neither of which are experienced by fluid particles.
\FloatBarrier
\begin{figure}
	\centering
	\vspace{-0mm}
	{\begin{overpic}
			[trim = 15mm 50mm 15mm 30mm,scale=0.8,clip,tics=20]{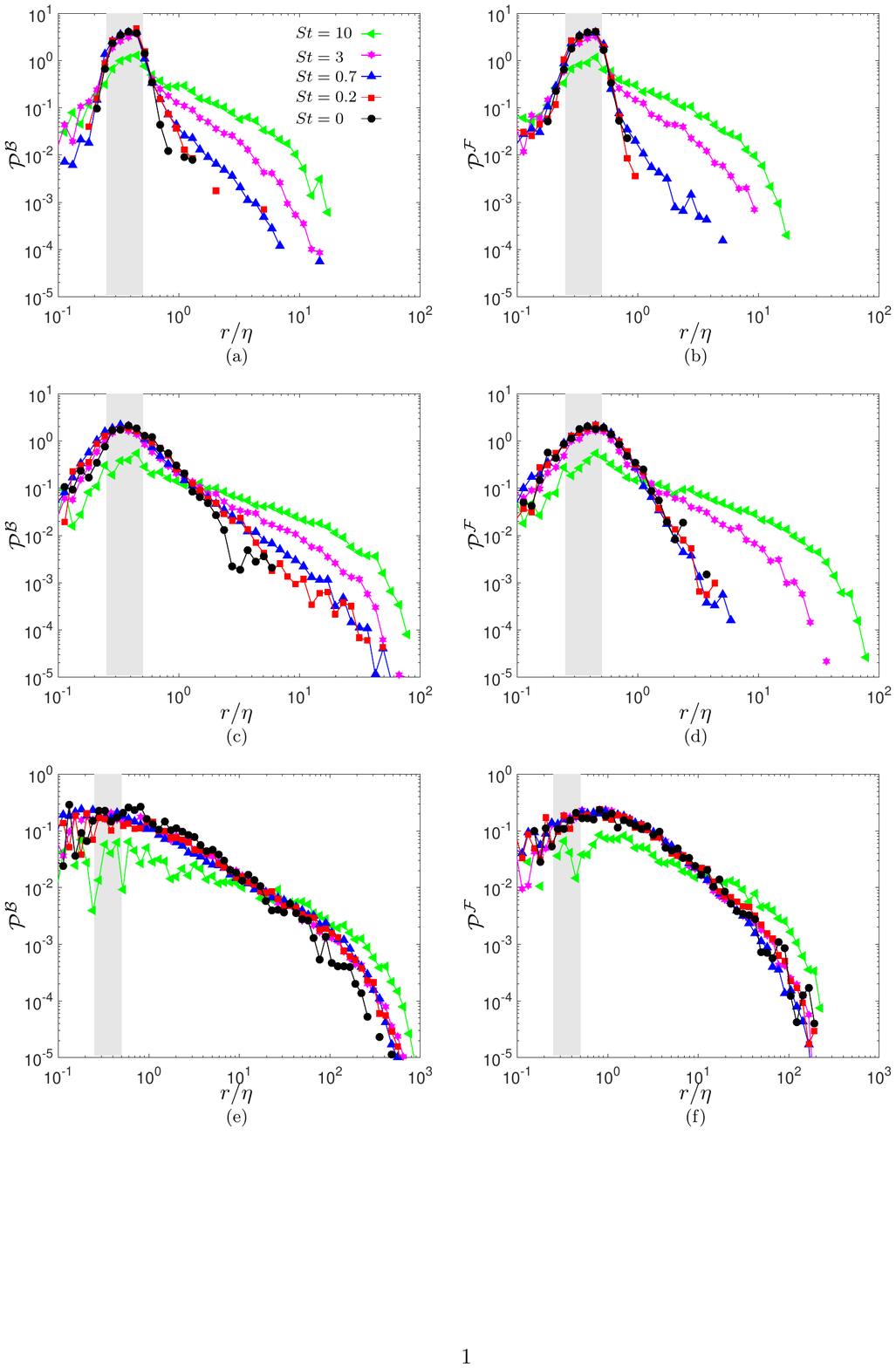}
	\end{overpic}}
	\caption{DNS results for $\mathcal{P^B}$ (plots (a),(c),(e)), $\mathcal{P^F}$ (plots (b),(d),(f)) with $\xi/\eta\in[0.25,0.5]$, various $St$ and (a),(b) $t=0.5\tau_\eta$, (c),(d) $t=2.5\tau_\eta$, (e),(f) $t=20\tau_\eta$.}
	\label{Ssep2}
\end{figure}
\FloatBarrier
\begin{figure}
\centering
{\begin{overpic}
[trim = 15mm 45mm 15mm 30mm,scale=0.8,clip,tics=20]{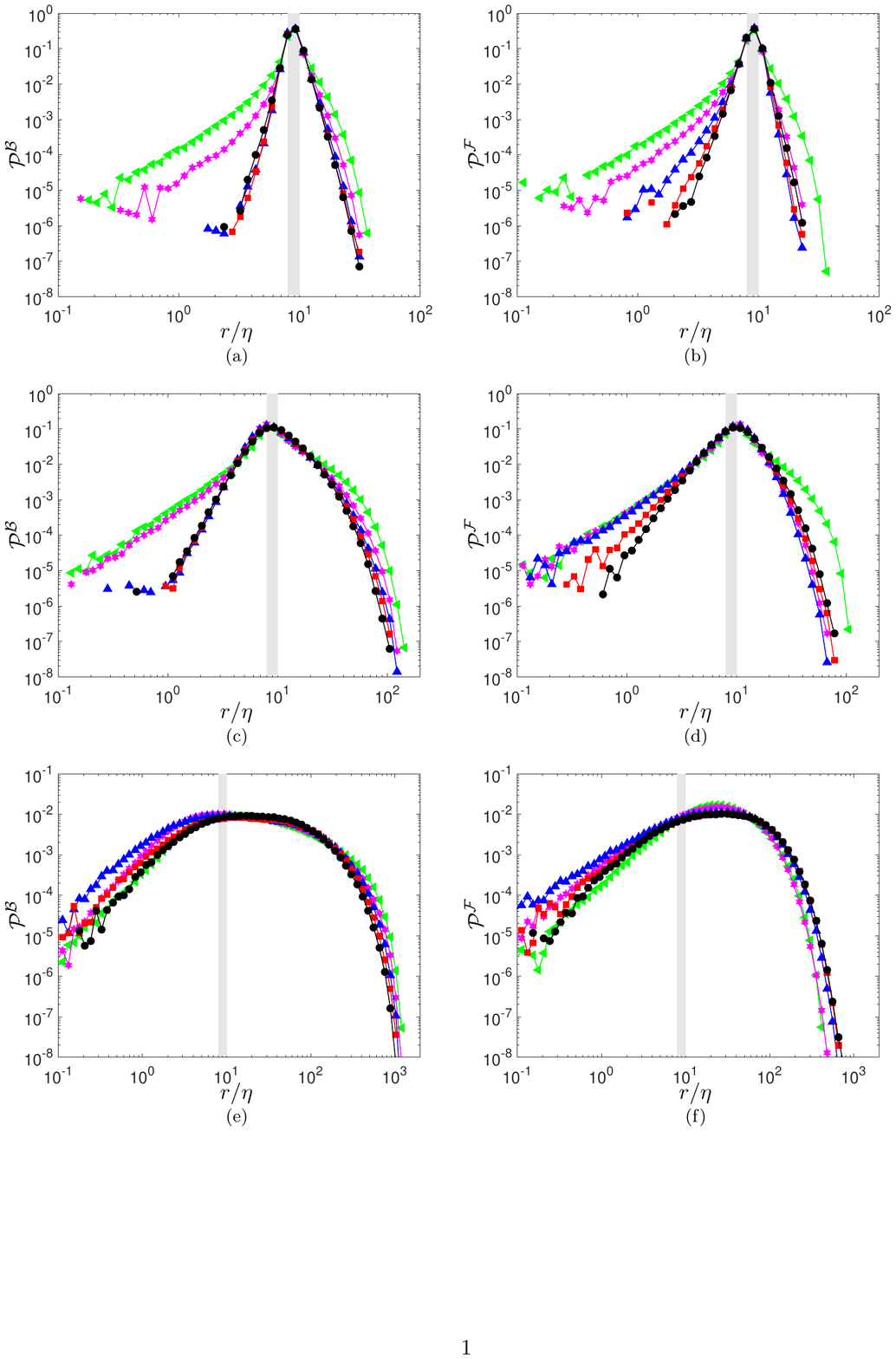}
\end{overpic}}
\caption{DNS results for $\mathcal{P}(r,-t\vert\xi,0)$ (plots (a),(c),(e)), $\mathcal{P}(r,t\vert\xi,0)$ (plots (b),(d),(f)) with $\xi/\eta\in[8,10]$, various $St$ and (a),(b) $t=0.5\tau_\eta$, (c),(d) $t=2.5\tau_\eta$, (e),(f) $t=20\tau_\eta$. Legend is the same as that in Figure~\ref{Ssep2}}
\label{Lsep2}
\end{figure}
\FloatBarrier
Having considered the irreversibility of the dispersion, we now consider the effect of $St$ on $\mathcal{P^B}$ and $\mathcal{P^F}$. The results in (\ref{Ssep2}) are for $\xi/\eta\in[0.25,0.5]$, and those in (\ref{Lsep2}) are for $\xi/\eta\in[8,10]$. In agreement with our arguments in \S\ref{theory}, for $\xi>\eta$, the results show that particle inertia affects the dispersion more strongly BIT than FIT, but the opposite for $r<\xi$. At short-times, inertia dramatically enhances the dispersion rate both FIT and BIT, which is associated with their strongly non-local dynamics at the small-scales, and the associated formation of caustics in their velocity distributions. At long-times, inertia enhances the BIT dispersion, but reduces the FIT dispersion. This was also observed in \citet{bragg16} for the mean-square separation, and the same explanation we gave there applies here: the opposite effect of inertia FIT and BIT at long-times is because of the opposite effect of the non-local contribution to the particle dynamics depending upon whether they are separating or approaching. 

The results also show that inertia changes the functional dependence of $\mathcal{P^B}$ and $\mathcal{P^F}$, except for large $s/\tau_\eta$ and $r\gg\xi$. This is due to the difference in the PDFs of $\bm{w}^p$ and $\Delta\bm{u}$, the difference being very strong in the dissipation range \citep{bec10a,ireland16a}. 

\begin{figure}
\centering
{\begin{overpic}
[trim = 15mm 135mm 15mm 0mm,scale=0.8,clip,tics=20]{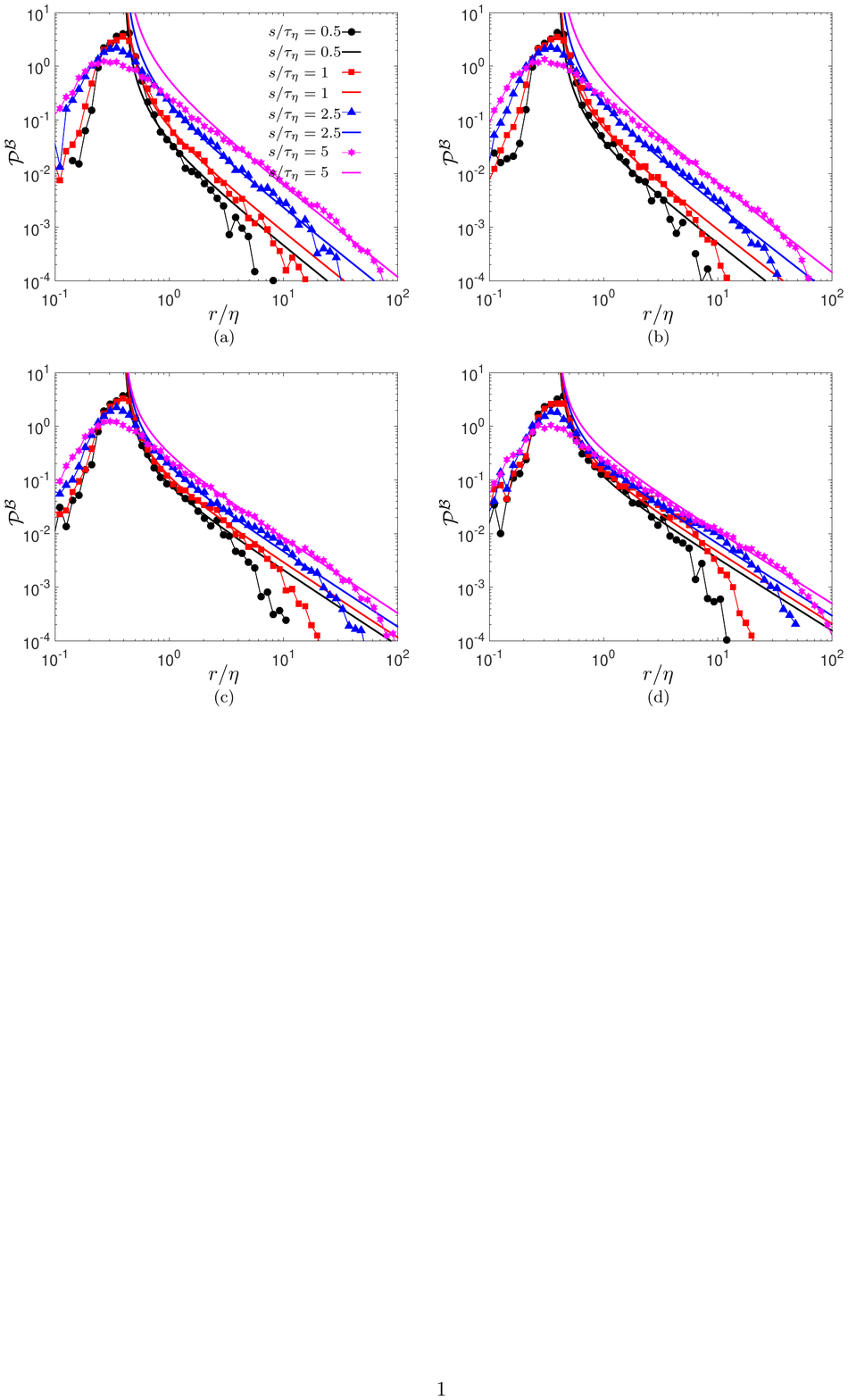}
\end{overpic}}
\caption{DNS results (symbols) and theoretical prediction (\ref{Pr4}) (lines) for $\mathcal{P^B}$ with $\xi/\eta\in[0.25,0.5]$ and (a) $St=0.7$, (b) $St=1$, (c) $St=2$, (d) $St=3$.}
\label{SsepTheory}
\end{figure}

\begin{figure}
	\centering
	{\begin{overpic}
			[trim = 15mm 135mm 15mm 0mm,scale=0.8,clip,tics=20]{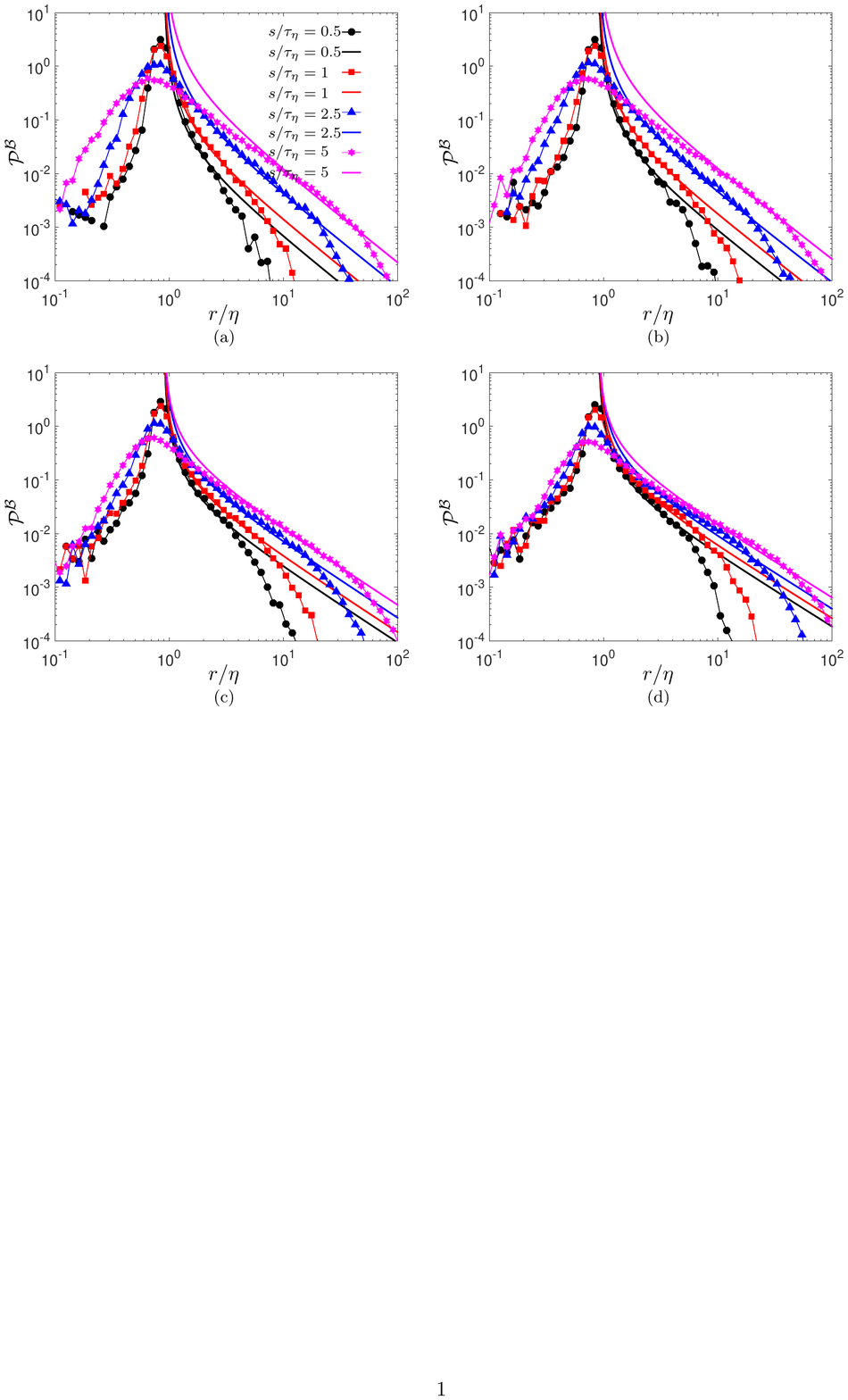}
	\end{overpic}}
	\caption{DNS results (symbols) and theoretical prediction (\ref{Pr4}) (lines) for $\mathcal{P^B}$ with $\xi/\eta\in[0.75,1]$ and (a) $St=0.7$, (b) $St=1$, (c) $St=2$, (d) $St=3$.}
	\label{SsepTheory2}
\end{figure}

We now test the theoretical prediction in (\ref{Pr3}) for $St\geq O(1)$ and $\xi$ in the dissipation range. As explained in \S\ref{TheorScal}, (\ref{Pr3}) does not fully predict the $s$ dependence, but only the $r$-dependence
\begin{align}
\mathcal{P^B}({r},-s\vert{\xi},0)\propto\frac{r}{\sqrt{3r^2-2\xi^2}}\Big\vert\xi- \sqrt{3r^2-2\xi^2}\Big\vert^{d_2-4},\label{Pr4}
\end{align}
which we compare with our DNS data in figure~\ref{SsepTheory} and figure~\ref{SsepTheory2}, using the DNS data for $d_2(St)$ (and in plotting (\ref{Pr4}) we used the center value of the DNS sets $\xi/\eta\in[0.25,0.5]$ and $\xi/\eta\in[0.75,1]$). 
\FloatBarrier
\FloatBarrier
The results show that (\ref{Pr4}) describes $\mathcal{P^B}$ for $r>\xi$ very well, capturing both the transitional scaling for $r=O(\xi)$ and small $s$, and also the scaling of the tails. At each $s$, (\ref{Pr4}) does not describe the scaling of the far tails of $\mathcal{P^B}$, and at longer times, (\ref{Pr4}) overpredicts $\mathcal{P^B}$ for small $r$ close to the peak. This is simply because these regimes are dominated by particles whose relative velocity does not correspond to the part of the relative velocity PDF described by (\ref{GM}). Indeed, (\ref{Pr4}) has to fail in the limit $r/\xi\to\infty$ in order for the moments of $\mathcal{P^B}$ to be finite. In Scatamacchia \emph{et al.} \citep{scatamacchia12}, they estimated that for $\mathcal{P^F}$ for fluid particles, the cut off scale $r_c(s)$, beyond which the separation $\mathcal{P^F}$ should be effectively zero, should be limited by the fluid r.m.s velocity $u'$, and the estimate $r_c(s)\approx u's$ described the DNS data well. For inertial particles and for $\mathcal{P^B}$, such a simple scaling does not work since figure~\ref{Ssep2} shows that $r_c(t)$ increases dramatically with increasing $St$, yet in isotropic turbulence, $v'\leq O(u')$ \citep{ireland16a}, where $v'$ is inertial particle r.m.s. velocity.
\begin{figure}
\centering
{\begin{overpic}
[trim = 15mm 200mm 15mm 0mm,scale=0.8,clip,tics=20]{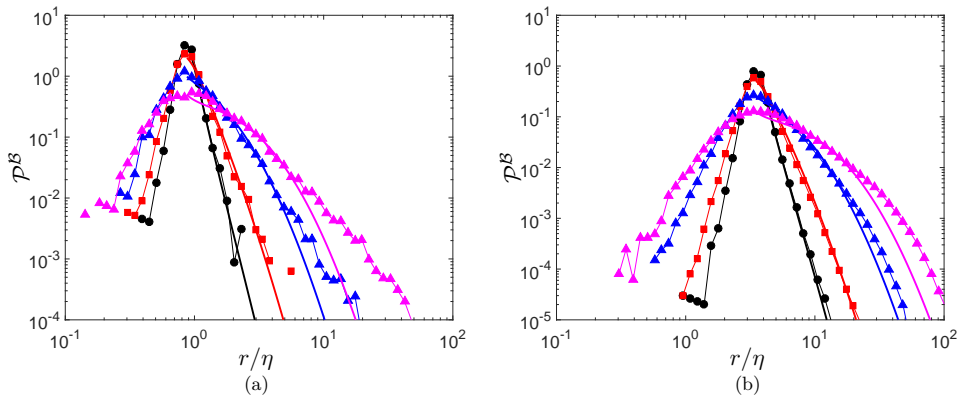}
\end{overpic}}
\caption{DNS results (symbols) and theoretical prediction (\ref{Pr2Newb2}) (lines) for $\mathcal{P^B}$ with (a) $\xi/\eta\in[0.75,1]$, (b) $\xi/\eta\in[3,4]$ and $St=0$. Legend is the same as that in Figure~\ref{SsepTheory2}.}
\label{St0plot}
\end{figure}
\FloatBarrier
In figure~\ref{St0plot} we consider the theoretical prediction for $\mathcal{P^B}$ and $St=0$ given in (\ref{Pr2Newb2}) (compared with the results in figure~\ref{SsepTheory}, the statistical noise in the data for $St=0$ at $\xi/\eta\in[0.25,0.5]$ is too strong, and so we plot $\xi/\eta\in[3,4]$ instead). The predictions of (\ref{Pr2Newb2}) are in excellent agreement with the DNS data for  $s\leq\tau_\eta$ and $r\geq\xi$. The discrepancies for $s>\tau_\eta$ are mainly due to the breakdown of (\ref{rfsol}). It is interesting that the theory breaks down first in describing the far tail of $\mathcal{P^B}$, corresponding to $r\gg\xi$. A possible explanation for this is that the timescales of the extreme values of $\Delta\bm{u}$ are shorter than those of small to moderate values of $\Delta\bm{u}$, and that as a consequence, the short-time expansion breaks down faster in describing the extreme events in the dispersion. For example, we can estimate the timescale of $\Delta\bm{u}$ conditioned on $\|\Delta\bm{u}\|$ as $\tau_{\Delta u}\sim r/\|\Delta\bm{u}({r},t)\|$, which shows that the correlation timescale of $\Delta\bm{u}$ decreases as $\|\Delta\bm{u}\|$ increases (for fixed separation).

\begin{figure}
\centering
{\begin{overpic}
[trim = 12mm 135mm 15mm 0mm,scale=0.8,clip,tics=20]{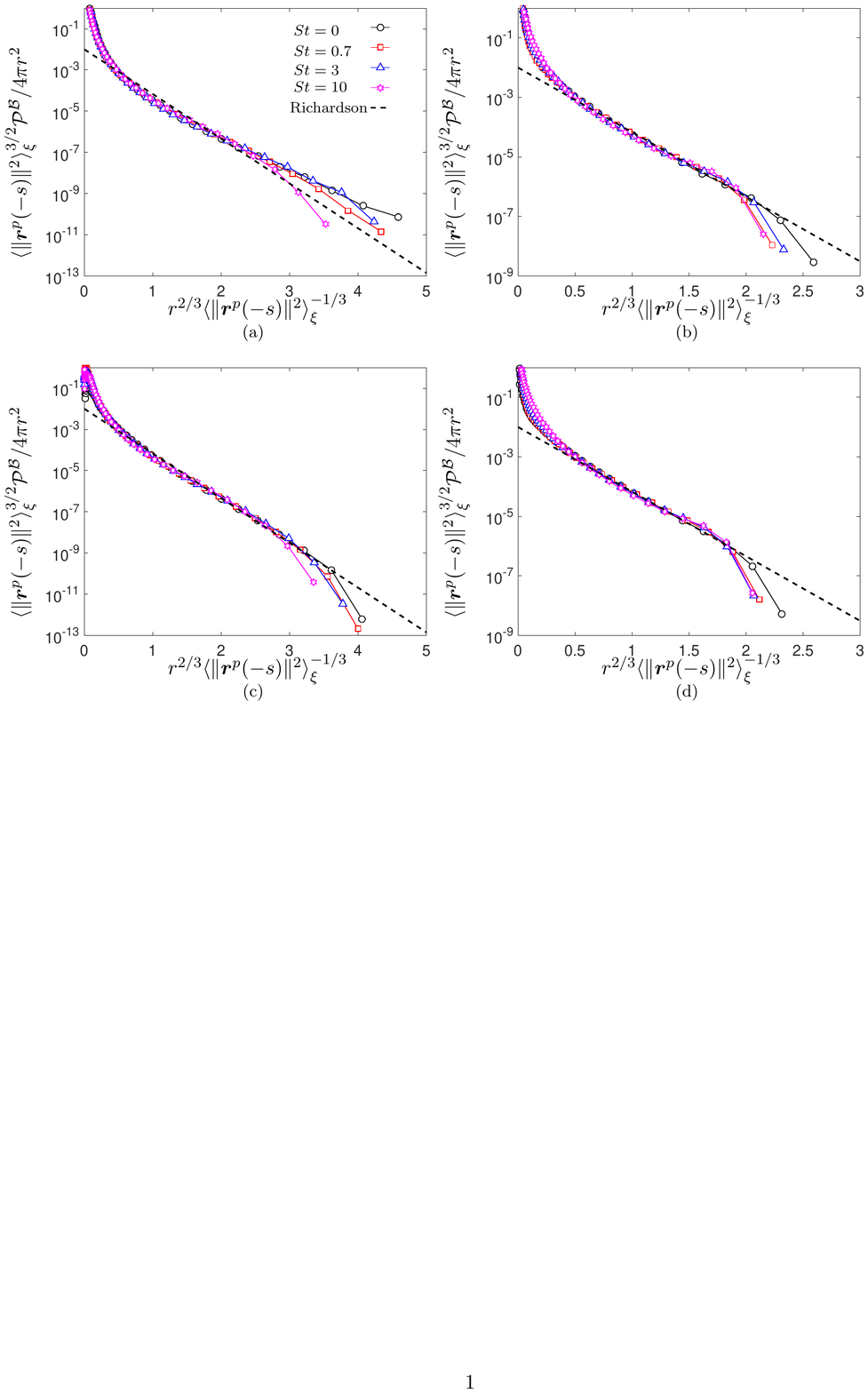}
\end{overpic}}
\caption{DNS results for $\mathcal{P^B}$ for (a),(b) $\xi/\eta\in[0,2]$, (c),(d) $\xi/\eta\in[8,10]$ and (a),(c) $s/\tau_\eta=20$, (b),(d) $s/\tau_\eta=80$.}
\label{Richplot}
\end{figure}

In figure~\ref{Richplot} we consider how $\mathcal{P^B}$ at long times compares with the Richardson PDF scaling. Based upon the idea that the non-local contribution to the inertial particle relative velocities is quite weak for $St_r\leq O(1)$ with $r$ in the inertial range, we argued that the inertial particle diffusion coefficient in the inertial range should scale with $r$ the same way as Richardson's law assumes, i.e. $\propto r^{4/3}$. As a consequence, we expect that for whatever range Richardson's form of the dispersion PDF applies for $St=0$, it should also apply for inertial particles at long times. The results in figure~\ref{Richplot} confirm this expectation quite well, showing that Richardson's prediction for $\mathcal{P^B}$ holds for fluid particles and inertial particles over a similar range of separations. This is not trivially because $St_r\ll1$ in the inertial range. Indeed, for $St=10$, $St_r\geq O(1)$ for $r$ well into the inertial range in this DNS.
\FloatBarrier
The results in figure~\ref{Richplot} (a) and (b), which are for $\mathcal{P^B}$, are quite similar to those in Biferale \emph{et al.}  \citep{biferale14} for $\mathcal{P^F}$, although they were able to probe much further into the tails of the PDF than we are, due to the very large number of particles in their DNS. In both our results and theirs, deviations from Richardson's scaling is observed for small and large $r^{2/3}\langle\|\bm{r}^p(-s)\|^2\rangle_{\xi}^{-1/3}$. A natural explanation for this would be to attribute these deviations to the effects of the dissipation and integral scales of motion on the dispersion. However, Biferale \emph{et al.}  \citep{biferale14} re-computed $\mathcal{P^F}$ where only particle-pairs with separations in the range $r\in[25,300]\eta$ were included in the PDF (i.e. excluding pairs lying outside the inertial range), and the results showed that the strong departures in the right tail still remained, leading them to conclude that the deviations could not be caused by the influence of the large-scales. Unfortunately, due to computational limitations, we were not able to re-compute $\mathcal{P^B}$ for particles lying in the range $r\in[25,300]\eta$, but this is something that should be checked in future work to test for the influence of finite $Re_\lambda$ effects on $\mathcal{P^B}$  compared with its effect on $\mathcal{P^F}$. 
\begin{figure}
\centering
{\begin{overpic}
[trim = 15mm 135mm 15mm 0mm,scale=0.8,clip,tics=20]{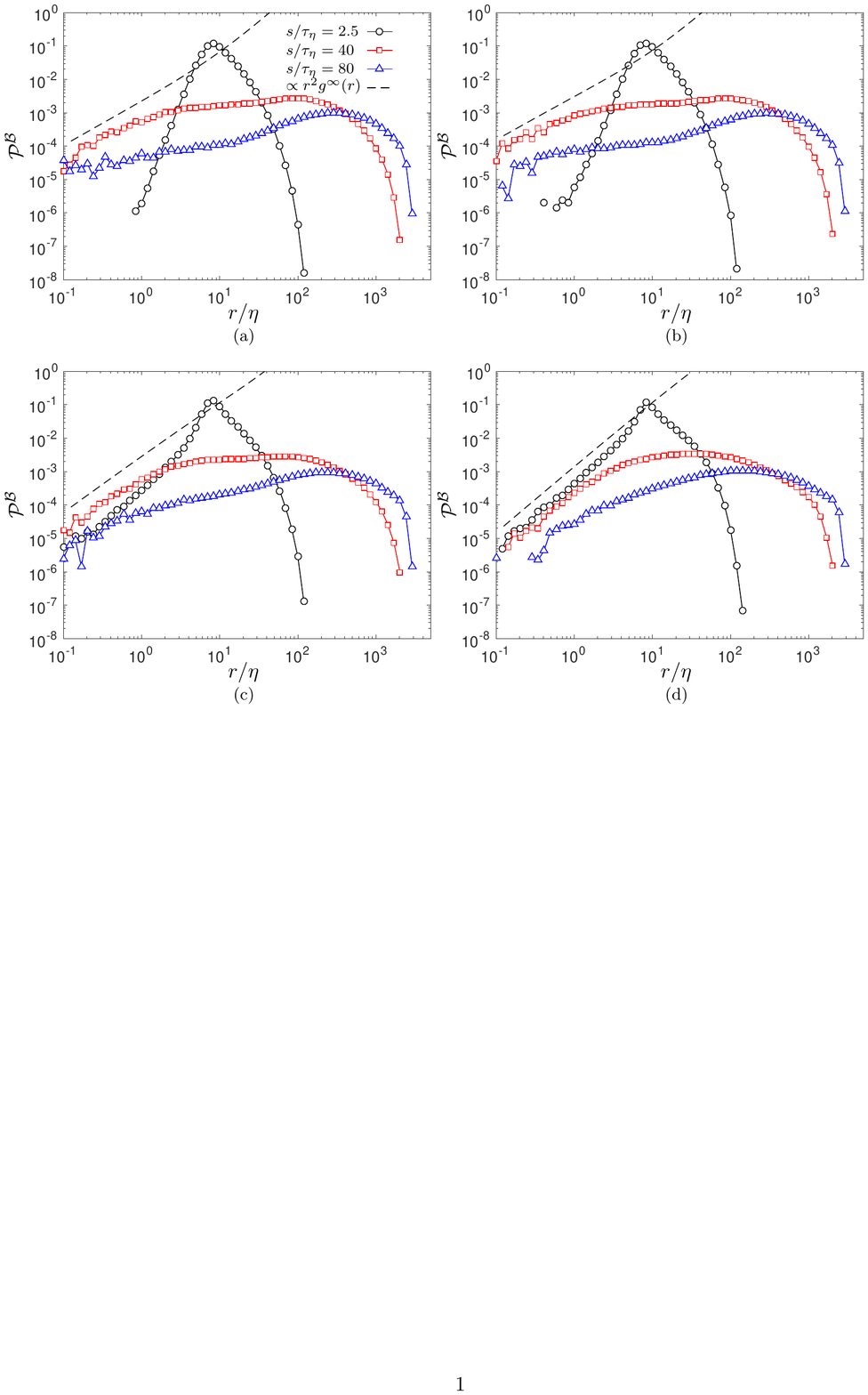}
\end{overpic}}
\caption{DNS results for $\mathcal{P^B}$ (using $\xi/\eta\in[8,10]$) and $r^2g^\infty(r)$ with (a) $St=0.5$, (b) $St=0.7$, (c) $St=3$, (d) $St=10$.}
\label{Lsep3}
\end{figure}
\FloatBarrier
Finally, in figure~\ref{Lsep3} we consider the $\mathcal{P^B}$ for $\xi/\eta\in[8,10]$, and consider how it evolves towards the $s/\tau^p_\xi\to\infty$ distribution, $\mathcal{P^B}(r,-s\to-\infty\vert\xi,0)\propto r^2 g^{\infty}(r)$. Over the time span we computed the PDFs in our DNS, this equilibrium solution is not realized. This is not surprising since one usually needs to wait several integral timescales before inertial particles in isotropic turbulence reach a steady state. What is interesting is that for intermediate $s$, $\mathcal{P^B}\propto r^2 g^{\infty}(r)$ for a period of time (and for $r$ in the dissipation range), before this scaling vanishes at larger $s$. A similar behavior was also observed in the DNS results by Biferale \emph{et al.} \citep{biferale14} and Bitane \emph{et al.} \citep{bitane13} for $\mathcal{P^F}$. For our BIT results the most likely explanation for this transitionary scaling is that at short-times, particles at $r<\xi$ correspond to particles that were at smaller separations in the past, and due to the small time scales in the dissipation range, their motion was in a qausi-equilibrium state. However, further back in time, the pairs will have passed through a minimum separation before which they were at larger separations. Consequently, as $s$ increases, more and more particles leave the dissipation range, and this gives rise to a scaling for $\mathcal{P^B}$ that varies with $r$ more weakly than $r^2 g^{\infty}(r)$. The same kind of argument would apply to $\mathcal{P^F}$.
\section{Conclusions}
In this paper we have analyzed, using theoretical and numerical methods, the forward in time (FIT) and backward in time (BIT) pair-separation PDFs for inertial particles in isotropic turbulence. In agreement with the earlier study by \citet{bragg16}, where the FIT and BIT mean-square separations were analyzed, we found that inertial particles separate much faster BIT than FIT, with the strength of the irreversibility depending upon the final/initial separation of the particle-pair and their Stokes number $St$. However, we also find that the irreversibility shows up in subtle ways in the behavior of the full PDF that it does not in the mean-square separation. For example, the irreversibility mechanisms affect the left and right tails of the FIT and BIT PDFs in different ways. Compared with fluid particles, inertia enhances the probability of finding particles with large separations at long times BIT, but reduces it FIT.

In the theory, we derived a prediction for the BIT/FIT PDF for $St\geq O(1)$, and for final/initial separations in the dissipation regime. The prediction reveals how caustics in the particle velocities in the dissipation range give rise to pair-separation PDFs with algebraically decaying tails. This result is universal, in the sense that it does not not depend upon the level of intermittency in the underlying turbulence. Comparisons with the DNS data showed that the prediction describes the pair-separation PDFs very well over its expected range of applicability, capturing the transitional behavior at small separations, where there is a strong effect of the final/initial separation. We also derived a prediction for the pair-separation PDF for fluid particles at short-times. In general, the result is given by a weighted integral of functions that depend upon the local flow topology. However, we argued that the integral is dominated by the contributions where the local fluid velocity field is undergoing strong extensional straining, and we used the multifractal formalism to describe the PDF of the fluid relative velocities. The resulting form of the PDF was shown to describe the DNS data very well for times up to the Kolmogorov timescale. 

The theoretical predictions for $St=0$ and $St\geq O(1)$ do not describe the left tail of the BIT/FIT PDFs, corresponding to particles that came from/go to smaller separation. Correctly predicting this left tail is more complex due to certain kinematic constraints on the trajectories in the short-time regime. Solving this problem is something that must be addressed in future work.

One of the important observations following from the theory and DNS results is that for $St\geq O(1)$, the extreme events in the pair-separation process do not arise, fundamentally, from the extreme events in the underlying dissipation range turbulence. For example, the probability of finding $St\geq O(1)$ particles with very large separations at short times is much greater than for fluid particles, and such intermittency in the dispersion of $St\geq O(1)$ particles would exist even in a flow where the fluid velocity field was Gaussian. The reason is that the extreme separation events for $St\geq O(1)$ particles are dominated by their non-local in-time dynamics and the associated phenomena of caustics, and not by the local, intermittent turbulent velocity field. One of the manifestations of this distinction is that our theoretical results predict that for final/initial separations in the dissipation range, the right tail of the BIT PDF decays more slowly with increasing $Re_\lambda$, whereas the opposite is true for inertial particles with $St\gtrsim 1$. We leave the testing of this prediction to future work.

At longer times, we argued that the Richardson PDF shape, that applies to fluid particles in the inertial range, should also apply to inertial particles with $St_r=O(1)$, where $St_r$ is the Stokes number based upon the eddy turnover timescale at separation $r$. This follows from the observation that in the inertial range, the non-local contribution to the inertial particle relative velocities is weak compared with the filtering and preferential sampling effects of inertia when $St_r\leq O(1)$, and that as a result, their diffusion coefficient scales like the Richardson prediction, namely $\propto r^{4/3}$. The DNS results confirmed this expectation, showing that over the range of $r$ for which Richardson's PDF shape holds for fluid particles, it also holds for inertial particles. The deviations from the Richardson shape at very large separations is most likely due to intermittency in the fluid turbulence, as was argued for the case of fluid particles by \citet{biferale14}.

\section*{Acknowledgements}

We gratefully acknowledge Peter J. Ireland for providing the DNS data used in this paper. We also gratefully acknowledge the anonymous reviewers of this paper for their thoughtful comments and questions, which have improved the quality of the paper.

\bibliographystyle{jfm}
\bibliography{refs_co12}

\end{document}